\documentclass[11pt]{article}

\usepackage[final]{acl}
\usepackage{booktabs}
\usepackage{multirow}
\usepackage{graphicx} 
\usepackage{times}
\usepackage{latexsym}

\usepackage[T1]{fontenc}

\usepackage[utf8]{inputenc}

\usepackage{microtype}

\usepackage{inconsolata}
\usepackage{graphicx}
\usepackage{amsmath} 
\usepackage{amssymb}
\usepackage{booktabs,xcolor,colortbl}
\usepackage{xcolor}
\definecolor{ctxblue}{HTML}{1F77B4}
\definecolor{ctlorange}{HTML}{FF7F0E}
\definecolor{decision}{HTML}{D62728} 
\definecolor{keepgreen}{HTML}{2E8B57}
\definecolor{tokenteal}{HTML}{0F766E}
\definecolor{tokengray}{HTML}{6B7280}
\usepackage{listings}
\usepackage{tcolorbox}
\usepackage{fontawesome5}
\usepackage{xcolor}
\definecolor{darkblue}{RGB}{0,70,140}
\tcbuselibrary{skins,breakable,listings}

\usepackage[linesnumbered,ruled,vlined]{algorithm2e}

\tcbset{
  capboxstyle/.style={
    on line,
    tcbox raise base,
    nobeforeafter,
    boxrule=0.55pt,
    arc=2.2pt,
    boxsep=0.7pt,
    left=1.2pt,
    right=1.2pt,
    top=0.8pt,
    bottom=0.8pt,
    fontupper=\ttfamily\bfseries\footnotesize
  }
}

\newtcbox{\capbox}{capboxstyle,
  colframe=ctxblue!75!black,
  colback=ctxblue!6!white
}
\newtcbox{\capboxgreen}{capboxstyle,
  colframe=keepgreen!85!black,
  colback=keepgreen!8!white
}
\newtcbox{\capboxred}{capboxstyle,
  colframe=decision!80!black,
  colback=decision!7!white
}
\newtcbox{\capboxorange}{capboxstyle,
  colframe=ctlorange!85!black,
  colback=ctlorange!10!white
}
\newtcbox{\capboxteal}{capboxstyle,
  colframe=tokenteal!85!black,
  colback=tokenteal!8!white
}
\newtcbox{\capboxgray}{capboxstyle,
  colframe=tokengray!80!black,
  colback=tokengray!8!white
}

\DeclareRobustCommand{\tokinline}[1]{\ifmmode\text{#1}\else#1\fi}
\DeclareRobustCommand{\tok}[1]{\tokinline{\capbox{\detokenize{#1}}}}
\DeclareRobustCommand{\tokgreen}[1]{\tokinline{\capboxgreen{\detokenize{#1}}}}
\DeclareRobustCommand{\tokred}[1]{\tokinline{\capboxred{\detokenize{#1}}}}
\DeclareRobustCommand{\tokorange}[1]{\tokinline{\capboxorange{\detokenize{#1}}}}
\DeclareRobustCommand{\tokteal}[1]{\tokinline{\capboxteal{\detokenize{#1}}}}
\DeclareRobustCommand{\tokgray}[1]{\tokinline{\capboxgray{\detokenize{#1}}}}

\DeclareRobustCommand{\tokkeep}{\tokgreen{<KEEP>}}
\DeclareRobustCommand{\tokdrop}{\tokred{<DROP>}}
\DeclareRobustCommand{\tokneed}{\tokorange{<NEED>}}
\DeclareRobustCommand{\tokmid}{\tok{<MID>}}
\DeclareRobustCommand{\tokdone}{\tokgray{<DONE>}}
\DeclareRobustCommand{\tokselect}{\tokteal{<SELECT>}}
\DeclareRobustCommand{\tokpfx}{\tokteal{<PFX>}}
\DeclareRobustCommand{\toksfx}{\tokteal{<SFX>}}
\DeclareRobustCommand{\tokc}[1]{\tokteal{<C_#1>}}
\DeclareRobustCommand{\tokcend}[1]{\tokteal{</C_#1>}}

\newtcblisting{casecode}[2][]{%
  enhanced,
  listing only,
  listing engine=listings,
  boxrule=0.35pt,
  arc=1.6mm,
  outer arc=1.6mm,
  left=1.0mm,
  right=1.0mm,
  top=1.4mm,
  bottom=1.0mm,
  colback=white,
  colframe=black!18,
  before skip=0pt,
  after skip=0pt,
  title={#2},
  coltitle=black,
  fonttitle=\sffamily\bfseries\footnotesize,
  title filled=false,
  attach boxed title to top left={xshift=1.5mm,yshift*=-\tcboxedtitleheight/2},
  boxed title style={
    enhanced,
    colback=white,
    colframe=black!8,
    boxrule=0.25pt,
    arc=1.0mm,
    left=0.8mm,
    right=0.8mm,
    top=0.45mm,
    bottom=0.35mm
  },
  listing options={
    language=Python,
    basicstyle=\ttfamily\footnotesize,
    columns=fullflexible,
    keepspaces=true,
    upquote=true,
    showstringspaces=false,
    breaklines=true,
    breakatwhitespace=false,
    aboveskip=0pt,
    belowskip=0pt,
    frame=none,
    xleftmargin=0pt,
    xrightmargin=0pt,
    lineskip=1.5pt
  },
  #1
}

\tcbset{
  caseaccentteal/.style={
    colback=tokenteal!3!white,
    colframe=tokenteal!24!black,
    borderline west={1.3pt}{0pt}{tokenteal!72!black}
  },
  caseaccentkeep/.style={
    colback=keepgreen!4!white,
    colframe=keepgreen!24!black,
    borderline west={1.3pt}{0pt}{keepgreen!78!black}
  },
  caseaccentdrop/.style={
    colback=decision!3!white,
    colframe=decision!22!black,
    borderline west={1.3pt}{0pt}{decision!74!black}
  },
  caseaccentgray/.style={
    colback=tokengray!4!white,
    colframe=tokengray!22!black,
    borderline west={1.3pt}{0pt}{tokengray!72!black}
  },
  caseaccentblue/.style={
    colback=ctxblue!3!white,
    colframe=ctxblue!22!black,
    borderline west={1.3pt}{0pt}{ctxblue!72!black}
  },
  casecompact/.style={
    left=0.75mm,
    right=0.75mm,
    top=0.9mm,
    bottom=0.7mm,
    fonttitle=\sffamily\bfseries\scriptsize,
    attach boxed title to top left={xshift=1.2mm,yshift*=-\tcboxedtitleheight/2},
    boxed title style={
      enhanced,
      colback=white,
      colframe=black!8,
      boxrule=0.2pt,
      arc=0.9mm,
      left=0.6mm,
      right=0.6mm,
      top=0.3mm,
      bottom=0.2mm
    },
    listing options={
      language=Python,
      basicstyle=\ttfamily\scriptsize,
      columns=fullflexible,
      keepspaces=true,
      upquote=true,
      showstringspaces=false,
      breaklines=true,
      breakatwhitespace=false,
      aboveskip=0pt,
      belowskip=0pt,
      frame=none,
      xleftmargin=0pt,
      xrightmargin=0pt,
      lineskip=1.0pt
    }
  }
}

\newcommand{\g}[1]{{\small \shortstack[c]{+#1}}}

\title{\textsc{RepoShapley}: Shapley-Enhanced Context Filtering for Repository-Level Code Completion}

\author{
  \textbf{Yu Huo\textsuperscript{$\spadesuit$}\thanks{Equal contribution}},
  \textbf{Kun Zeng\textsuperscript{$\clubsuit$}\footnotemark[1]},
  \textbf{Siyu Zhang\textsuperscript{$\heartsuit$}},
  \textbf{Yuquan Lu\textsuperscript{$\clubsuit$}},
  \textbf{Cheng Yang\textsuperscript{$\diamondsuit$}},
  \textbf{Yifu Guo\textsuperscript{$\clubsuit$}},\\
  \textbf{Xiaoying Tang\textsuperscript{$\spadesuit$}\textsuperscript{$\sharp$}\textsuperscript{$\ddagger$}\thanks{Corresponding author}}
  \\
  \textsuperscript{$\spadesuit$}School of Science and Engineering, The Chinese University of Hong Kong, Shenzhen\\[-0.45ex]
  \textsuperscript{$\sharp$}Shenzhen Future Network of Intelligence Institute (FNii-Shenzhen)\\[-0.45ex]
  \textsuperscript{$\ddagger$}Guangdong Provincial Key Laboratory of Future Networks of Intelligence, CUHK(SZ)\\
  \textsuperscript{$\heartsuit$}University of California, San Diego\quad
  \textsuperscript{$\clubsuit$}Sun Yat-sen University\quad
  \textsuperscript{$\diamondsuit$}Hangzhou Dianzi University\\[-0.1ex]
  \makebox[\textwidth][c]{\small
    \href{mailto:yuhuo@link.cuhk.edu.cn}{\textcolor{darkblue}{\faIcon{envelope}}}\kern0.35em
    \textbf{Email:}\kern0.35em
    \href{mailto:yuhuo@link.cuhk.edu.cn}{\texttt{yuhuo@link.cuhk.edu.cn}},
    \href{mailto:tangxiaoying@cuhk.edu.cn}{\texttt{tangxiaoying@cuhk.edu.cn}}
  }\\[0.4ex]
  \makebox[\textwidth][c]{\small
    \href{https://github.com/yuhuo03/RepoShapley}{\textcolor{darkblue}{\faIcon{github}}}\kern0.35em
    \textbf{GitHub:}\kern0.35em
    \url{https://github.com/yuhuo03/RepoShapley}}
}

\hypersetup{
  pdftitle={RepoShapley: Shapley-Enhanced Context Filtering for Repository-Level Code Completion},
  pdfauthor={Yu Huo, Kun Zeng, Siyu Zhang, Yuquan Lu, Cheng Yang, Yifu Guo, Xiaoying Tang}
}

\begin{document}
\maketitle

\begin{abstract}
Repository-level code completion benefits from retrieval-augmented generation (RAG). However, controlling cross-file evidence is difficult because chunk utility is often interaction-dependent: some snippets help only when paired with complementary context, while others harm decoding when they conflict. We propose \textsc{RepoShapley}, a coalition-aware context filtering framework supervised by Shapley-style marginal contributions. Our offline labeling module, \textbf{ChunkShapley}, estimates signed per-chunk effects via teacher-forced probing, feeds them into a lightweight surrogate game that captures saturation and interference, computes exact Shapley values for small retrieval sets, and selects a decoding-optimal coalition through bounded post-verification with the frozen generator. The verified \tokkeep/\tokdrop\ decisions and retrieval triggers are then distilled into a single model via discrete control tokens. Experiments across benchmarks and backbones show that \textsc{RepoShapley} improves completion quality while reducing harmful context and unnecessary retrieval. 
\end{abstract}

\section{Introduction}

Large language models have demonstrated strong reasoning, coding, and generation capabilities~\cite{brown2020language,wei2022chain,chen2021evaluating,liu2026adaptivepromptstructurefactorization}. Yet repository-level code completion must resolve non-local dependencies such as project-specific APIs, shared contracts, and invariants~\cite{jimenez2024swe,ding2024cocomic}. Retrieval-Augmented Generation (RAG) injects cross-file evidence into Code LMs~\cite{lewis2020RAG,kang2024crag,shrivastava2023repofusion,bairi2023codeplan}, but retrieval control remains difficult under fixed context budgets: the model must filter redundant or misleading chunks from a noisy candidate pool~\cite{DingNIPS2024SemCoder,zhang2023repocoder,wei2025instructrag,liu2024lost,yoran2024making}.

\begin{figure}[t]
    \centering
    \includegraphics[width=\linewidth]{pictures/Result_Compare.png}
    \caption{Performance radar charts on StarCoder-Base-7B and CodeLlama-13B. The plots display relative improvements over the No-Retrieve baseline (center). \textsc{RepoShapley} achieves the best performance among compared methods across 11 tested metrics; see Table~\ref{tab:main_results}.}

    \label{fig:Performance}
\end{figure}

The core difficulty is that chunk utility is often interaction-dependent. A snippet may appear uninformative in isolation yet become decisive when paired with complementary context, such as an interface declaration together with its implementation. Conversely, a plausible chunk can degrade generation when it co-occurs with conflicting evidence, such as deprecated versus updated APIs~\cite{shi2023large,xu2024recomp}. Therefore, methods that score candidates independently can misestimate the utility of the multi-chunk context that is actually consumed at test time~\cite{Khandelwal2020Generalization,yan2024corrective,bertsch2025context}.

\begin{figure*}[t]
  \includegraphics[width=\linewidth]{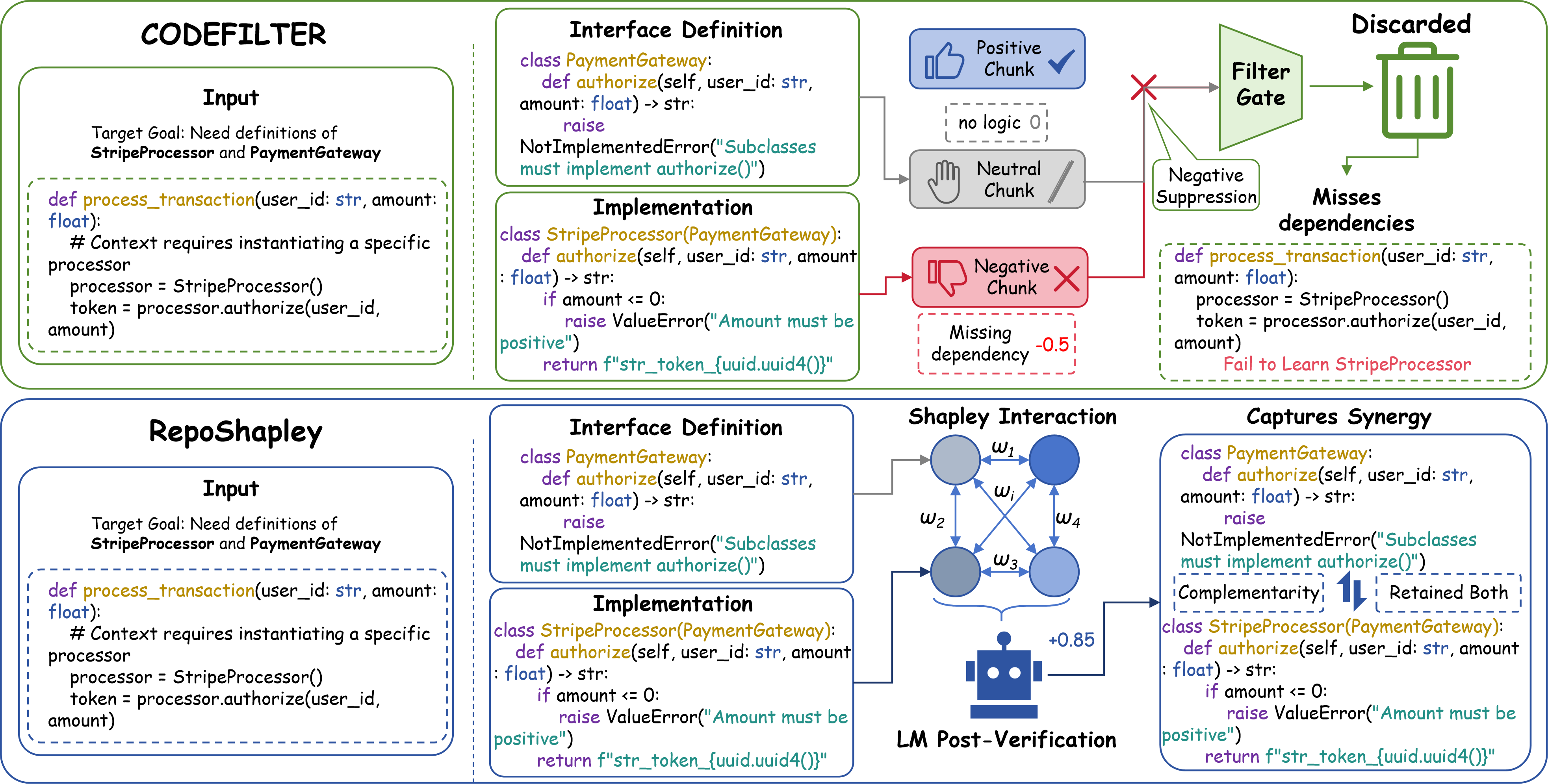}
  \caption{Under the same input context and the exact same retrieved candidate chunks, \texttt{CODEFILTER} makes decisions from independent per-chunk signals and can break under interaction effects, whereas \textsc{RepoShapley} performs coalition-aware filtering that more reliably removes high-score noise while preserving complementary evidence.}
  \label{fig:Compare}
\end{figure*}

To address this, we adopt a coalition-first approach. Retrieval control should be supervised by signals that reflect how a chunk behaves within a set, rather than in isolation. We introduce \textsc{RepoShapley}, a framework that learns to filter context using Shapley-style marginal contributions.

Our approach has two stages. First, we propose \textbf{ChunkShapley}, an offline labeling pipeline for interaction-aware supervision. Considering that computing Shapley values directly with the generator is prohibitive, we introduce a structured logistic surrogate that can capture saturation and conflict efficiently. We then apply a verification step to ground the selected coalitions in the generator's actual decoding behavior. Second, we distill the resulting coalition-derived labels into a single generator via discrete control tokens, which we call \textsc{RepoShapley}. This distillation enables efficient, interaction-aware retrieval control at inference time. As shown in Figure~\ref{fig:Performance}, \textsc{RepoShapley} achieves the best performance under our evaluated settings, supporting our motivation that coalition-aware supervision is crucial for difficult cross-file completion. Our contributions are as follows:

\begin{itemize}
    \item \textbf{Coalition-aware supervision for context filtering.}
    We formulate context selection as a cooperative game and use Shapley marginal contributions to capture complementarity and conflict beyond independent scoring.

    \item \textbf{ChunkShapley: Practical Shapley labeling for chunk filtering.}
    We combine single-chunk probing with a structured surrogate utility to compute exact Shapley values on small retrieval sets ($K{=}10$). We further select a verified coalition from a bounded candidate pool under decoding-time metrics.

    \item \textbf{\textsc{RepoShapley}: distillation for online retrieval control.}
    We distill verified keep and drop decisions into discrete control tokens, enabling a single model to decide when to retrieve and which chunks to keep.
\end{itemize}


\section{Related Work}

\paragraph{Repository-Level RAG and Retrieval Control.}
RAG mitigates non-local dependencies in code completion by retrieving cross-file evidence~\cite{lewis2020RAG,izacard2021RAG,parvez2021retrieval,guu2020retrieval,jiang2023active,mallen2023not,yao2024adaptive}. Recent work improves context quality via iterative retrieval~\cite{gao2023retrieval,zhang2023repocoder,shrivastava2023repofusion,zhang2025coderag}, structure-aware indexing, including dataflow or call graphs~\cite{cheng2024dataflow,liu2024graphcoder}, and dedicated benchmarks~\cite{ding2023crosscodeeval,liu2024repobench,li2025codefilter,generation2024coderag,yang2025deep}. In parallel, retrieval control has received increasing attention, focusing on when to retrieve and what to retain under a fixed context budget. RepoFormer~\cite{Wu2024repoformer} triggers retrieval through self-evaluation, while \texttt{CODEFILTER}~\cite{li2025codefilter} filters chunks using independent likelihood-based signals. However, these controllers largely assess chunks in isolation. As a result, they do not explicitly account for combinatorial interactions such as complementarity between interfaces and implementation. In contrast, we cast context filtering as a coalition scoring problem to model such inter-dependencies.

\paragraph{Shapley Values in RAG and Supervision.}
Shapley values~\cite{shapley1953value} provide an axiomatic notion of marginal contribution and have been widely used in interpretability, including SHAP-style formulations~\cite{Lundberg2017shapley,ghorbani2019data,sundararajan2017axiomatic}. In RAG, prior work applies Shapley-style analysis to attribute outputs to retrieved documents~\cite{nematov2025source,ye2025fair} or to estimate token-level importance~\cite{asai2024self,xiao2025tokenshapley}. Our use differs along three axes. First, whereas SHAP and Data Shapley~\cite{ghorbani2019data} perform \emph{post-hoc} attribution on a frozen model, we use Shapley marginalization to \emph{construct supervision} for an active retrieval controller. Second, document-level Shapley in RAG~\cite{nematov2025source,ye2025fair} treats each retrieved passage as an independent player; we instead operate at the \emph{chunk} level within a single repository and explicitly model coalition effects such as saturation and conflict. Third, TokenShapley~\cite{xiao2025tokenshapley} attributes importance to individual tokens, whereas our formulation attributes importance to \emph{subsets} of retrieved chunks, capturing inter-chunk synergies that token-level analysis cannot express. We then distill the resulting coalition reasoning into a token-level policy, enabling practical retrieval decisions during generation.

\section{Methodology}
\subsection{Repository-level Retrieval-Augmented Code Completion}

Repository-level code completion requires grounding generation in cross-file information such as project-specific APIs, shared utilities, and type or contract conventions. RAG addresses this by retrieving candidate snippets from the repository. However, retrieved evidence is often interaction-heavy: a snippet may be useful only when paired with complementary context, and seemingly relevant snippets can degrade generation when they introduce conflicting implementations.

\paragraph{Problem setup.}
Given a repository $\mathcal{R}$ and a target file, each instance is represented as $(X_{\text{in}}, X_{\text{out}}, Y)$.
Here $X_{\text{in}}=(X_p,X_s)$ is the in-file context in fill-in-the-middle (FIM) format with prefix $X_p$ and suffix $X_s$,
$X_{\text{out}}$ denotes a cross-file pool constructed from other files in $\mathcal{R}$,
and $Y$ is the ground-truth missing span between $X_p$ and $X_s$ \cite{zhang2023repocoder,Wu2024repoformer}.

\paragraph{Retrieval and generation.}
A retriever $R$ queries $X_{\text{out}}$ with $X_{\text{in}}$ and returns top-$K$ candidate chunks
$X_{\text{cc}} = R(X_{\text{in}}, X_{\text{out}})=\{cc_1,\ldots,cc_K\}$.
A generator $G_{\theta}$ then predicts the completion $\hat{Y}$ conditioned on $X_{\text{in}}$ and a selected subset $X_S \subseteq X_{\text{cc}}$.
Hence, the key problem is to estimate chunk utility and retain the subset that best supports generating $Y$.

\begin{figure*}[t]
  \includegraphics[width=\linewidth]{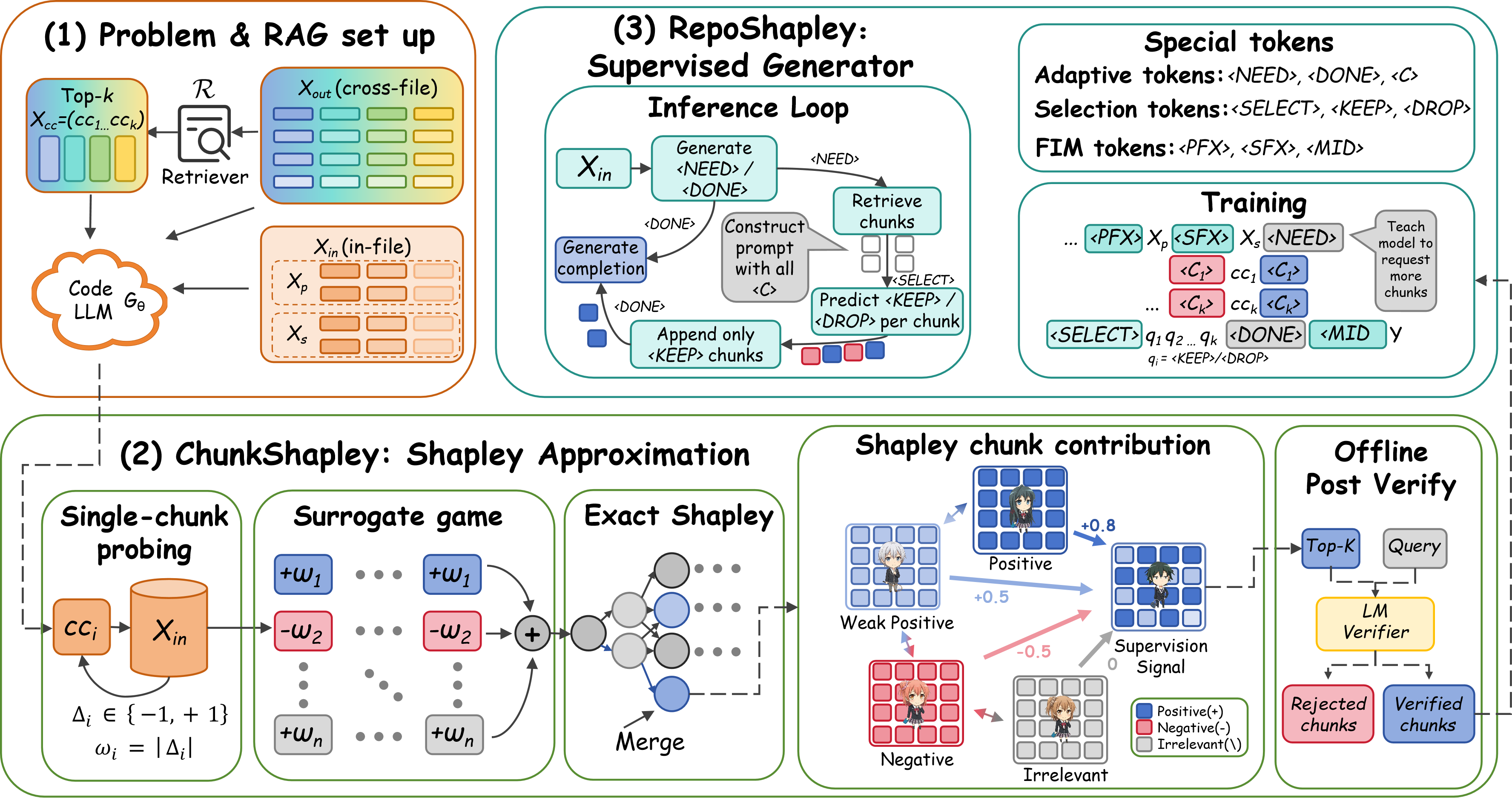}
  \caption{The overall framework of \textsc{RepoShapley}. The pipeline consists of two phases: (2) An offline ChunkShapley module that estimates the interaction-aware contribution of each chunk; and (3) An online Shapley-supervised Generator trained to control retrieval and filter contexts based on the estimated Shapley values.}
  \label{fig:Framework}
\end{figure*}

\subsection{Interaction-aware Chunk Attribution via Shapley Values}

\paragraph{Why independent chunk scoring is insufficient.}
Retrieved code snippets rarely contribute independently. A chunk can be uninformative on its own but become essential when paired with complementary context such as an interface and its implementation. Conversely, a seemingly relevant snippet may reduce generation quality when it conflicts with other retrieved evidence. As a result, per-chunk scores computed in isolation can be a poor proxy for the utility of the multi-chunk context used at test time.

\paragraph{Subset utility as a cooperative game.}
We therefore evaluate chunks at the set level. Given top-$K$ candidates, we treat each chunk as a player and any subset as a coalition. Let $D=\{1,\ldots,K\}$ index candidates and $S\subseteq D$ denote a coalition, with $X_S=\{cc_i: i\in S\}$. We define the coalition value as the normalized teacher-forced log-likelihood gain on the ground-truth completion:

\begin{align*}
\label{eq:coalition_value}
v(S \mid X_{\text{in}}, Y)
&= \ell(X_{\text{in}}, X_S) - \ell(X_{\text{in}}) \\
\ell(C)&=\frac{1}{|Y|}\log p_\theta(Y \mid C).
\end{align*}

where $\log p_{\theta}(Y \mid C)=\sum_{t=1}^{|Y|}\log p_{\theta}(y_t \mid y_{<t}, C)$. By construction, $v(\emptyset\mid X_{\text{in}},Y)=0$, and $v(S)$ can be negative when retrieved context decreases model likelihood.
Appendix~\ref{app:ablation_utility} compares log-likelihood with metric-based utilities (EM/ES) and shows log-likelihood yields the best downstream performance.

\paragraph{Shapley attribution.}
We quantify interaction-aware chunk contributions using the Shapley value~\cite{shapley1953value}, which is defined as the average marginal gain of chunk $i$ over all coalitions:

\[
\label{eq:shapley_subset}
\begin{aligned}
\phi_i
&= \sum_{S\subseteq D\setminus\{i\}}
\frac{|S|!\,(K-|S|-1)!}{K!}\,
\Delta v_i(S)\\
\Delta v_i(S)
&= v(S\cup\{i\}\mid X_{\text{in}},Y) - v(S\mid X_{\text{in}},Y).
\end{aligned}
\]

Intuitively, $\phi_i>0$ indicates that chunk $i$ is helpful on average across different co-occurring contexts, while $\phi_i\le 0$ suggests redundancy or harm under interactions.
Shapley values satisfy \emph{efficiency}: $\sum_{i\in D}\phi_i = v(D\mid X_{\text{in}},Y)$, allowing negative attributions when some chunks reduce coalition utility.

\subsection{ChunkShapley: Practical Shapley Labeling for Chunk Filtering}

Exact Shapley computation under the true coalition utility $v(\cdot)$ is impractical, as it would require evaluating the generator on exponentially many subsets.
We therefore propose \textbf{ChunkShapley}, an \emph{offline} labeling pipeline that (a) probes each chunk once to obtain a signed effect, (b) defines a lightweight surrogate game to approximate interaction patterns, (c) computes exact Shapley values under the surrogate by enumerating all $2^K$ coalitions, which is inexpensive since $v_{\text{sur}}(\cdot)$ is closed-form, and (d) performs bounded post-verification with the frozen generator to ground final keep and drop labels in decoding-time behavior. Algorithmic details are deferred to Appendix Algorithm~\ref{alg:chunkshapley}.

\paragraph{(a) Single-chunk probing.}
We first compute a per-instance baseline score using teacher forcing and probe each candidate in isolation.
Let $\ell(C)$ denote the normalized teacher-forced log-likelihood.
For each retrieved chunk $cc_i$, we define its single-chunk effect

\begin{align*}
\label{eq:delta_single}
\Delta_i &= \ell(X_{\text{in}},\{cc_i\}) - \ell(X_{\text{in}})\\
y_i &=\mathrm{sign}(\Delta_i),\quad \omega_i =|\Delta_i|.
\end{align*}

To ensure consistent likelihood estimation under a limited context window, we preserve the full target span $Y$ and apply \textbf{left-truncation} only to the input context (i.e., $X_{\text{in}}$ and retrieved chunks).

\paragraph{(b) Logistic surrogate game.}
While ranking by $\Delta_i$ captures individual relevance, it ignores coalition dynamics. To model interactions efficiently, we define a one-dimensional surrogate utility. Given $(y_i,\omega_i)$, we aggregate coalition $S$ via a weighted vote:

\begin{align*}
g(S)&=\sum_{i\in S}\omega_i y_i,\quad v_{\text{sur}}(S)=\sigma(\beta\, g(S))-\sigma(0).
\end{align*}

where $\sigma(\cdot)$ is the sigmoid and $\beta>0$ controls the saturation scale.
This surrogate is not meant to match the full combinatorial utility; it targets two dominant effects for filtering.
The sigmoid yields diminishing returns: when $|g(S)|$ is large, $\sigma'(\beta g(S))\approx 0$, so additional similarly-signed evidence contributes little, capturing redundancy under a fixed budget.
Conflicts are expressed by negative votes ($y_i=-1$), which reduce $g(S)$ and can suppress $v_{\text{sur}}(S)$ even when some chunks are individually helpful. Subtracting $\sigma(0)$ ensures $v_{\mathrm{sur}}(\varnothing) = 0$ and keeps utilities centered.
The surrogate remains lightweight for exhaustive subset evaluation, while any residual mismatch to decoding-time behavior is addressed by verification.

\paragraph{(c) Exact Shapley values under the surrogate.}
We compute Shapley values using the subset form under the surrogate utility:
\[
\phi_i
=
\frac{1}{K}
\sum_{S\subseteq D\setminus\{i\}}
\frac{v_{\text{sur}}(S\cup\{i\})-v_{\text{sur}}(S)}{\binom{K-1}{|S|}}.
\]
Since our $v_{\text{sur}}(S)$ is closed-form, evaluating all $2^K$ subsets is computationally negligible for small retrieval sizes ($K\le 10$). This allows us to obtain \emph{exact} Shapley values under $v_{\text{sur}}$, avoiding the variance of sampling approximations.

In contrast, computing interactions using the heavy generator $G_{\theta}$ would require exponentially many coalition evaluations and is intractable. Therefore, we use $\phi_i$ under the surrogate as a proposal signal and rely on post-verification to finalize the decision.

\paragraph{(d) Post-verification via a bounded candidate pool.}
Because decoding quality is non-monotonic in context, positive attributions alone do not guarantee improved greedy decoding. Since the surrogate is only a proxy, we verify a small candidate pool with the frozen generator and select the coalition that maximizes decoding-time quality. This step is used only for offline label construction with access to $Y$; inference never uses $Y$.

Let $\pi_\phi$ and $\pi_\Delta$ be indices sorted by $\phi_i$ and $\Delta_i$.
We build a de-duplicated set $\mathcal{C}$ containing:
(i) Shapley prefixes $\{\pi_\phi[1{:}n]\}_{n=1}^{N_v}$,
(ii) short $\Delta$ prefixes as a strong single-chunk baseline, and
(iii) size-2/3 combinations among top-$L$ chunks by $\Delta$ to explicitly probe local synergies.
For each $S\in\mathcal{C}$, we decode with the frozen generator and choose
\[
S^\star=\arg\max_{S\in\mathcal{C}}
\big(\mathrm{ES}(\hat{Y}_S,Y),\ \mathrm{EM}(\hat{Y}_S,Y)\big)
\]
using lexicographic maximization (ES first, EM as tie-break).
We then treat $S^\star$ as the teacher keep/drop labels for distillation.

\paragraph{Verified labels for retrieval triggering.}
The post-verification step also yields an oracle decision on whether retrieval is necessary.
Let $\hat{Y}_{\emptyset}$ be the decoding result using only in-file context $X_{\text{in}}$, and let
$\hat{Y}_{S^\star}$ be the decoding result using the verification-selected coalition $S^\star$.
Let $\Delta_{\mathrm{ES}}=\mathrm{ES}(\hat{Y}_{S^\star},Y)-\mathrm{ES}(\hat{Y}_{\emptyset},Y)$.
We define the retrieval-control label as
\[
\label{eq:rstar}
r^\star =
\begin{cases}
    \tokdone, & \text{if }\Delta_{\mathrm{ES}} \le \epsilon\\
    \tokneed, & \text{otherwise}.
\end{cases}
\]
where $\epsilon$ is a small margin tuned on the validation set (default $\epsilon=0$ unless stated otherwise).
This label is used only for offline supervision; inference never accesses $Y$.

\subsection{\textsc{RepoShapley}: Distilling ChunkShapley into Signal Tokens}
While ChunkShapley provides robust coalition-aware supervision, the pipeline is too computationally intensive for online use.
We therefore propose \textsc{RepoShapley}, which \emph{distills} verified coalition decisions into discrete control tokens,
enabling a single generator to efficiently decide \emph{when} to retrieve and \emph{which} chunks to retain at inference time.

\paragraph{Signal tokens and verified labels.}
We introduce retrieval-control tokens
$\mathcal{T}_R=\{\tokneed,\tokdone\}$
to decide whether cross-file evidence is required, and candidate-selection tokens
$\mathcal{T}_S=\{\tokkeep,\tokdrop\}$
to indicate which retrieved chunks should be retained.

Step~(d) outputs a \emph{verification-selected coalition} $S^\star$ by evaluating a small set of Shapley-proposed
candidate coalitions $\mathcal{C}$ using the frozen generator under decoding-time constraints, to match decoding-time behavior.
We treat $S^\star$ as the teacher keep/drop label set and distill it into token-level supervision by assigning, for each
retrieved chunk $cc_i$,
\[
Q(cc_i)=
\begin{cases}
\tokkeep & \text{if } i\in S^\star\\
\tokdrop & \text{otherwise}.
\end{cases}
\]
In this way, surrogate Shapley signals are used only to propose promising coalitions,
while the student model learns to imitate the \emph{verified} coalition-level behavior encoded by $S^\star$,
turning combinatorial subset selection into a single-shot, controllable generation policy at inference time.

\paragraph{Training: two-format verbalized supervision.}
Following the standard separation of evidence selection and completion generation in retrieval-augmented code modeling, we train a single model with two serialized views of each instance.
Format-1 supervises \emph{selection}: given the in-file context and retrieved candidates, the model emits a keep/drop decision token for each chunk.
Format-2 supervises \emph{generation}: the model produces the missing span conditioned only on the kept evidence.
Both formats reuse the same control tokens and share all parameters, enabling the model to learn selection and generation within a unified autoregressive interface.

\smallskip
\noindent\textbf{Format-1: Selection.}
Given the in-file context and the retrieved candidate list, the model predicts a length-$K$ decision sequence
$q_{1:K}\in\{\tokkeep,\tokdrop\}^K$ under a dedicated
\tokselect{} marker.
Let $[X_p]$ and $[X_s]$ denote tokenized FIM prefix and suffix, and let $\mathrm{Pack}(X_{\text{cc}})$ be the deterministic
serialization of retrieved candidates $X_{\text{cc}}=\{cc_1,\dots,cc_K\}$:
\[
\begin{aligned}
\mathrm{Pack}(X_{\text{cc}})= {} & \tokc{1}[cc_1]\tokcend{1}\ \cdots \\
& \tokc{K}[cc_K]\tokcend{K}.
\end{aligned}
\]
The Format-1 sequence is
\[
\begin{aligned}
\mathsf{F1}:\quad
&\tokpfx [X_p]\ \toksfx [X_s]\ \tokneed\ \\
& \mathrm{Pack}(X_{\text{cc}})\
\tokselect\ q_1\ q_2\ \cdots\ q_K\ \tokdone.
\end{aligned}
\]

We supervise $q_i$ using the \emph{verified teacher coalition} $S^\star$:
$q_i^\star=\tokkeep$ if $i\in S^\star$ and $\tokdrop$ otherwise.

\smallskip
\noindent\textbf{Format-2: Generation.}
To teach the model how to complete code \emph{given filtered evidence}, we construct a generation format that includes
only the chunks in $S^\star$ and then decodes the target span in FIM mode:
\[
\begin{aligned}
\mathsf{F2}:\quad
&\tokpfx [X_p]\ \toksfx [X_s]\ \tokneed\ 
\mathrm{Pack}(C_{S^\star})\\
&\ \tokdone\ \tokmid\ [Y].
\end{aligned}
\]

\smallskip
\noindent\textbf{No-retrieval format.}
If retrieval is unnecessary ($r^\star=\tokdone$), we drop the cross-file block and the selection head:
\[
\tokpfx [X_p]\ \toksfx [X_s]\ \tokdone\ 
\tokmid\ [Y].
\]
This indicates that in-file context suffices.

\smallskip
\noindent\emph{Remark.}
We reuse \tokdone{} both as the retrieval decision token and as a block delimiter; the two usages are unambiguous from their fixed positions in the sequence.
The retrieval-control token (\tokneed/\tokdone) is learned with teacher forcing as a next-token target (counted in $\mathcal{L}_R$), rather than provided as an oracle input.

\begin{algorithm}[t]
\SetAlgoLined
\DontPrintSemicolon
\caption{\textsc{RepoShapley} Inference Process}
\label{alg:repshapley_infer_batch}
\KwIn{
    Generator $G$, Retriever $R$, Cross-file pool $X_{\text{out}}$, In-file context $X_{\text{in}}=(X_p,X_s)$;\\
    Token sets $\mathcal{T}_R=\{\tokneed,\tokdone\}$,
    $\mathcal{T}_S=\{\tokkeep,\tokdrop\}$; threshold $t_c$.
}
\KwOut{Completed code $\hat{Y}$.}

$X \leftarrow (\tokpfx, X_p, \toksfx, X_s)$\; 

$r \leftarrow \mathrm{Select}(\mathrm{Softmax}_{\mathcal{T}_R}(G(\cdot\mid X)), t_c)$\;

\If{$r = \tokdone$}{
    $X \leftarrow \mathrm{append}(X, \tokmid)$\;
    \Return $\hat{Y} \leftarrow G(X)$\;
}

$X_{\text{cc}} \leftarrow R(X_{\text{in}}, X_{\text{out}})$\; 

$X_{\text{sel}} \leftarrow X \oplus \tokneed \oplus \mathrm{Pack}(X_{\text{cc}})
\oplus \tokselect$\;

$(q_1,\dots,q_K) \leftarrow G(X_{\text{sel}})$\; 

$\hat{S} \leftarrow \{i \in \{1,\dots,K\}: q_i=\tokkeep\}$\;

$X \leftarrow X \oplus \tokneed \oplus \mathrm{Pack}(C_{\hat{S}})\oplus \tokdone$\;

$X \leftarrow \mathrm{append}(X, \tokmid)$\;

\Return $\hat{Y} \leftarrow G(X)$\;
\end{algorithm}

\begin{table*}[t]
\centering
\setlength{\tabcolsep}{4pt}
\renewcommand{\arraystretch}{0.9}

\caption{Code completion performance in the Infilling setting.}
\vspace{-5pt}

\resizebox{\textwidth}{!}{%
\begin{tabular}{l|l| cccccc |ccc |cc}
\toprule
\multirow{3}{*}{\textbf{Model}} & \multirow{3}{*}{\textbf{Strategy}} & \multicolumn{6}{c}{\textbf{RepoEval}} & \multicolumn{3}{c}{\textbf{CCLongEval}} & \multicolumn{2}{c}{\textbf{CCEval}} \\
\cmidrule(lr){3-8} \cmidrule(lr){9-11} \cmidrule(lr){12-13}
 & & \multicolumn{2}{c}{\textbf{Line}} & \multicolumn{2}{c}{\textbf{API}} & \multicolumn{2}{c}{\textbf{Function}} & \multicolumn{2}{c}{\textbf{Chunk}} & \textbf{Func} & \multicolumn{2}{c}{\textbf{Line}} \\
\cmidrule(lr){3-4} \cmidrule(lr){5-6} \cmidrule(lr){7-8} \cmidrule(lr){9-10} \cmidrule(lr){11-11} \cmidrule(lr){12-13}
 & & EM (M1) & ES (M2) & EM (M3) & ES (M4) & UT (M5) & ES (M6) & EM (M7) & ES (M8) & ES (M9) & EM (M10) & ES (M11) \\ \midrule

\multirow{5}{*}{\textbf{SC-Base-1B}} 
& No-Retrieve & 43.14 & 67.39 & 38.03 & 66.81 & 21.67 & 47.29 & 30.62 & 60.54 & 47.16 & 18.72 & 42.85 \\
& Full-Retrieve & 52.27 & 73.13 & 44.18 & 69.09 & 25.61 & 55.93 & 37.49 & 64.04 & 50.72 & 22.38 & 47.26 \\
& RepoFormer & 54.71 & 76.52 & 45.73 & 72.41 & 28.46 & 57.69 & 41.93 & 70.21 & 54.37 & 25.42 & 49.18 \\
& \texttt{CODEFILTER} & 57.19 & 78.84 & 48.37 & 75.66 & 31.13 & 59.91 & 44.52 & 72.48 & 56.59 & 27.81 & 52.03 \\
\rowcolor{gray!20}\cellcolor{white} & \textsc{RepoShapley} & \textbf{61.34}\g{4.15} & \textbf{82.78}\g{3.94} & \textbf{53.62}\g{5.25} & \textbf{79.53}\g{3.87} & \textbf{35.84}\g{4.71} & \textbf{64.39}\g{4.48} & \textbf{48.57}\g{4.05} & \textbf{77.52}\g{5.04} & \textbf{61.18}\g{4.59} & \textbf{32.26}\g{4.45} & \textbf{56.37}\g{4.34} \\ \midrule

\multirow{5}{*}{\textbf{SC-Base-3B}} 
& No-Retrieve & 48.12 & 72.38 & 40.17 & 68.91 & 24.93 & 51.52 & 36.16 & 65.19 & 49.63 & 21.82 & 45.58 \\
& Full-Retrieve & 57.84 & 77.21 & 48.83 & 72.68 & 30.58 & 58.16 & 42.61 & 68.29 & 53.84 & 25.92 & 50.31 \\
& RepoFormer & 58.59 & 79.16 & 49.82 & 74.63 & 32.89 & 60.62 & 46.38 & 72.11 & 56.39 & 28.85 & 52.16 \\
& \texttt{CODEFILTER} & 61.21 & 81.09 & 51.97 & 77.62 & 35.18 & 63.26 & 49.62 & 74.58 & 58.51 & 30.84 & 55.29 \\
\rowcolor{gray!20}\cellcolor{white} & \textsc{RepoShapley} & \textbf{64.93}\g{3.72} & \textbf{85.27}\g{4.18} & \textbf{56.38}\g{4.41} & \textbf{81.72}\g{4.10} & \textbf{39.91}\g{4.73} & \textbf{68.16}\g{4.90} & \textbf{53.52}\g{3.90} & \textbf{78.83}\g{4.25} & \textbf{62.84}\g{4.33} & \textbf{35.79}\g{4.95} & \textbf{59.41}\g{4.12} \\ \midrule

\multirow{5}{*}{\textbf{SC-Base-7B}} 
& No-Retrieve & 51.62 & 75.51 & 43.83 & 71.29 & 25.62 & 52.71 & 38.91 & 66.62 & 52.84 & 23.37 & 48.01 \\
& Full-Retrieve & 58.26 & 77.79 & 50.38 & 75.01 & 32.26 & 60.21 & 44.62 & 69.19 & 55.16 & 28.51 & 52.91 \\
& RepoFormer & 59.83 & 79.26 & 51.31 & 77.46 & 35.71 & 61.19 & 46.84 & 74.16 & 57.11 & 29.62 & 55.49 \\
& \texttt{CODEFILTER} & 61.49 & 81.41 & 53.62 & 79.29 & 37.79 & 63.41 & 49.16 & 77.26 & 59.84 & 32.11 & 57.84 \\
\rowcolor{gray!20}\cellcolor{white} & \textsc{RepoShapley} & \textbf{65.81}\g{4.32} & \textbf{86.59}\g{5.18} & \textbf{58.79}\g{5.17} & \textbf{84.11}\g{4.82} & \textbf{41.84}\g{4.05} & \textbf{68.16}\g{4.75} & \textbf{54.73}\g{5.57} & \textbf{81.29}\g{4.03} & \textbf{64.62}\g{4.78} & \textbf{36.59}\g{4.48} & \textbf{62.91}\g{5.07} \\ \midrule

\multirow{5}{*}{\textbf{Llama-7B}} 
& No-Retrieve & 51.89 & 73.42 & 41.53 & 66.98 & 24.81 & 44.56 & 37.21 & 65.16 & 50.37 & 18.16 & 43.34 \\
& Full-Retrieve & 60.18 & 78.91 & 48.76 & 73.16 & 29.93 & 52.21 & 45.41 & 69.37 & 52.11 & 23.41 & 47.46 \\
& RepoFormer & 60.52 & 79.36 & 49.31 & 75.91 & 33.19 & 52.64 & 46.84 & 69.56 & 52.16 & 24.26 & 48.31 \\
& \texttt{CODEFILTER} & 63.76 & 82.31 & 52.62 & 78.54 & 32.84 & 54.49 & 50.76 & 74.91 & 54.74 & 27.16 & 51.68 \\
\rowcolor{gray!20}\cellcolor{white} & \textsc{RepoShapley} & \textbf{68.31}\g{4.55} & \textbf{86.76}\g{4.45} & \textbf{57.28}\g{4.66} & \textbf{83.11}\g{4.57} & \textbf{37.56}\g{4.72} & \textbf{59.14}\g{4.65} & \textbf{55.21}\g{4.45} & \textbf{79.19}\g{4.28} & \textbf{59.41}\g{4.67} & \textbf{31.23}\g{4.07} & \textbf{55.24}\g{3.56} \\ \midrule

\multirow{5}{*}{\textbf{Llama-13B}} 
& No-Retrieve & 53.81 & 74.84 & 42.19 & 67.96 & 26.31 & 47.14 & 41.91 & 67.46 & 52.71 & 20.97 & 45.88 \\
& Full-Retrieve & 61.41 & 79.29 & 49.81 & 77.41 & 31.69 & 54.21 & 47.36 & 70.61 & 55.24 & 25.53 & 50.20 \\
& RepoFormer & 62.19 & 81.51 & 50.46 & 79.21 & 34.39 & 54.44 & 48.96 & 71.11 & 55.41 & 27.20 & 52.20 \\
& \texttt{CODEFILTER} & 64.16 & 82.71 & 53.01 & 78.99 & 35.41 & 57.76 & 51.31 & 74.19 & 58.81 & 29.91 & 54.69 \\
\rowcolor{gray!20}\cellcolor{white} & \textsc{RepoShapley} & \textbf{68.89}\g{4.73} & \textbf{87.11}\g{4.40} & \textbf{57.66}\g{4.65} & \textbf{83.41}\g{4.42} & \textbf{40.11}\g{4.70} & \textbf{62.39}\g{4.63} & \textbf{55.91}\g{4.60} & \textbf{78.59}\g{4.40} & \textbf{63.51}\g{4.70} & \textbf{35.05}\g{5.14} & \textbf{59.37}\g{4.68} \\ \bottomrule
\end{tabular}%
}
\label{tab:main_results}
\end{table*}

\paragraph{Objectives with masked contexts.}
We mask all in-file and cross-file \emph{content} tokens in the loss and compute gradients only on \emph{generated targets}
(control tokens, selection tokens, and the completion $Y$).
Let $r^\star\in\mathcal{T}_R=\{\tokneed,\tokdone\}$ be the retrieval-control label.
For retrieval-needed instances (Formats $\mathsf{F1}/\mathsf{F2}$), $r^\star=\tokneed$;
for no-retrieval instances, $r^\star=\tokdone$.
For Format-1, we optimize retrieval triggering and selection:
\[
\begin{aligned}
\mathcal{L}^{\mathsf{F1}}_R &= - \log P_G\!\big(r^\star \mid X_{\text{in}}\big)\\
\mathcal{L}^{\mathsf{F1}}_S &= - \sum_{i\in\mathcal{J}}\log P_G\!\big(q_i^\star \mid X_{\text{in}}, X_{\text{cc}}, r^\star, q_{<i}^\star\big)\\
\mathcal{L}^{\mathsf{F1}} &= \lambda_R\mathcal{L}^{\mathsf{F1}}_R+\lambda_S\mathcal{L}^{\mathsf{F1}}_S
\end{aligned}
\]
where $\mathcal{J}=\{1,\dots,K\}$ for retrieval-needed instances and $\mathcal{J}=\emptyset$ for no-retrieval instances.

For Format-2, we optimize retrieval triggering and generation conditioned on the verified filtered context.
Let $X_{S^\star}=\mathrm{Pack}(C_{S^\star})$ denote the serialized filtered evidence.
\[
\begin{aligned}
\mathcal{L}^{\mathsf{F2}}_R &= - \log P_G\!\big(r^\star \mid X_{\text{in}}\big)\\
\mathcal{L}^{\mathsf{F2}}_Y &= - \sum_{t=1}^{T} \log P_G\!\big(y_t \mid y_{<t}, X_{\text{in}}, X_{S^\star}, r^\star\big)\\
\mathcal{L}^{\mathsf{F2}} &= \lambda_R\mathcal{L}^{\mathsf{F2}}_R+\mathcal{L}^{\mathsf{F2}}_Y.
\end{aligned}
\]
Here $\mathcal{L}_R$ is implemented as the cross-entropy on the next-token prediction at the retrieval-control position
(i.e., immediately after $\toksfx [X_s]$), rather than a separate classifier.

During training, we either (i) include both formats for each instance, or (ii) sample one format per instance with a fixed mixing ratio.
The final objective is the expectation over the chosen format:
\[
\mathcal{L}=\mathbb{E}_{\mathsf{F}\sim \pi}\big[\mathcal{L}^{\mathsf{F}}\big],
\quad \mathsf{F}\in\{\mathsf{F1},\mathsf{F2}\}.
\]

\paragraph{Inference.}
At inference time, \textsc{RepoShapley} makes retrieval decisions in one autoregressive rollout.
Given the in-file context, the model first predicts a retrieval-control token
$r\in\mathcal{T}_R={\tokneed,\tokdone}$.
If $r=\tokdone$, it directly performs FIM decoding to generate the completion.

If $r=\tokneed$, we retrieve $K$ cross-file candidates
$X_{\text{cc}}={cc_1,\dots,cc_K}$ and serialize them as $\mathrm{Pack}(X_{\text{cc}})$.
Conditioned on this packed block, the model outputs a length-$K$ selection sequence under
\tokselect, where $(q_1,\dots,q_K)\in\mathcal{T}_S^K$ and
$\mathcal{T}_S={\tokkeep,\tokdrop}$.
We then keep only chunks with $q_i=\tokkeep$, append them to the prompt, and generate $\hat{Y}$ via FIM decoding after emitting $\tokmid$.
Alg.~\ref{alg:repshapley_infer_batch} provides the full procedure. We use $\oplus$ to denote token sequence concatenation.

\section{Experiments}

\subsection{Experimental Setup}
\paragraph{Dataset.}
We curate 290k Python repositories from The Stack \cite{kocetkov2023stack} after strict quality filtering (LOC constraints, AST parsing, and deduplication; Appendix~\ref{sec:appendix_a}).
Following \cite{Wu2024repoformer}, we sample 7.5k repositories to construct 50k labeled instances: for each instance, we retrieve top-10 cross-file chunks using Jaccard similarity \cite{jaccard1912distribution} and assign supervision derived from ChunkShapley. During data labeling, we discard instances whose verification-selected coalition $S^\star$ fails to reach a minimum completion quality, i.e., $\mathrm{ES}(\hat{Y}_{S^\star},Y) < \tau_{\text{es}}$, to ensure supervision reliability.
We split repositories into disjoint 95\%/5\% train/validation pools before instance construction, so validation instances come from repositories unseen during training.

\paragraph{Models and Training.}
We fine-tune StarCoderBase (SCB-1B/3B/7B) \cite{li2023starcoder} and CodeLlama (Llama-7B/13B) \cite{roziere2023codellama} for 2 epochs using a learning rate of $2\times10^{-5}$ with linear decay and 5\% warm-up.
We set $\lambda_R=\lambda_S=2.0$, max sequence length to 4096. With a global batch size of 512 on 8 NVIDIA H100 (80GB), training takes on average 2.2/6.5/15.4 hours for SCB-1B/3B/7B and 15.8/28.6 hours for Llama-7B/13B, respectively. Details are shown in Appendix~\ref{app:hparam}.

\paragraph{Benchmarks and Metrics.}
We evaluate on three repository-level code completion benchmarks:
RepoEval~\cite{zhang2023repocoder}, CrossCodeEval~\cite{ding2023crosscodeeval}, and CrossCodeLongEval~\cite{Wu2024repoformer}.
Together they cover line, API, chunk, and function-level completion tasks under realistic cross-file dependencies.
We consider two prompting settings: \textbf{Infilling} (FIM with $X_{\text{in}}=(X_p,X_s)$) and \textbf{Left-to-right} (prefix-only with $X_{\text{in}}=X_p$).
Following prior work~\cite{Wu2024repoformer}, we report Exact Match (EM) and Edit Similarity (ES) for non-function tasks, and unit-test pass rate (UT) for function tasks. Metric formulations are shown in Appendix~\ref{metrics}.

\paragraph{Baselines.}
We compare \textsc{RepoShapley} against:
(1) \textbf{No-Retrieve} (in-file only);
(2) \textbf{Full-Retrieve} \cite{zhang2023repocoder} (top-10 sparse retrieval);
(3) \textbf{RepoFormer} \cite{Wu2024repoformer} (selective retrieval); and
(4) \textbf{\texttt{CODEFILTER}} \cite{li2025codefilter} (likelihood-based filtering).
\texttt{CODEFILTER} serves as the primary baseline to highlight the benefit of interaction-aware supervision.

\begin{table}[t]
\centering
\setlength{\tabcolsep}{1.5pt} 
\renewcommand{\arraystretch}{1.}
\caption{Component Ablation of \textsc{RepoShapley} on RepoEval. We investigate components in (A) Labeling and (B) Distillation. Baseline is SC-Base-1B.}
\label{tab:ablation_component}

\resizebox{\linewidth}{!}{%
\scriptsize
\begin{tabular}{l|cc|cc|c}
\toprule
\multirow{2}{*}{\textbf{Method / Variant}} & \multicolumn{2}{c|}{\textbf{RepoEval-Line}} & \multicolumn{2}{c|}{\textbf{RepoEval-API}} & \textbf{Latency} \\
\cmidrule(lr){2-3} \cmidrule(lr){4-5}
 & \textbf{EM} & \textbf{ES} & \textbf{EM} & \textbf{ES} & (ms/req) \\
\midrule
RepoFormer & 54.71 & 76.52 & 45.73 & 72.41 & 661 \\
\texttt{CODEFILTER} & 57.19 & 78.84 & 48.37 & 75.66 & 947 \\
\rowcolor{gray!15} 
\textbf{\textsc{RepoShapley}} & \textbf{61.34} & \textbf{82.78} & \textbf{53.62} & \textbf{79.53} & 1053 \\ 
\midrule

\multicolumn{6}{l}{\textit{\textbf{A. Labeling Strategy}}} \\
1. w/o Post-verification & 38.50 & 54.44 & 36.15 & 55.81 & -- \\
2. $\Delta$-only labeling & 58.45 & 77.12 & 48.46 & 75.26 & -- \\
3. Linear Surrogate & 59.92 & 76.41 & 50.73 & 77.09 & -- \\
4. Uniform Weights & 60.18 & 80.97 & 51.82 & 77.38 & -- \\
\midrule

\multicolumn{6}{l}{\textit{\textbf{B. Distillation}}} \\
5. Format-1 only & 5.56 & 8.27 & 2.34 & 5.66 & 523 \\
6. Format-2 only & 59.88 & 79.11 & 52.12 & 77.49 & 830 \\
7. No Trigger & 61.26 & 81.33 & 52.15 & 78.81 & 1462 \\

\bottomrule
\end{tabular}%
}
\vspace{-5pt}
\end{table}

\subsection{Main Results}
Tables~\ref{tab:main_results} and \ref{tab:left_to_right_results} show that \textsc{RepoShapley} consistently improves repository-level infilling across benchmarks and backbones, validating our core hypothesis that supervision derived from evidence coalitions better reflects interaction-heavy retrieval.

First, interaction-blind filtering remains brittle. While adaptive controllers generally outperform Full-Retrieve, methods trained from per-chunk labels (\texttt{CODEFILTER}) can still overfit to isolated similarity and fail to account for complementarity and conflict that only appear when multiple chunks are concatenated. This gap is most visible on harder settings that require resolving non-local dependencies such as Function, where selecting the right combination of evidence matters more.

Second, coalition-aware supervision yields the strongest gains on difficult tasks. On SC-Base-7B, \textsc{RepoShapley} improves RepoEval API from 53.62/79.29 to \textbf{58.79}/\textbf{84.11} (EM/ES) and raises RepoEval Function unit-test pass rate from 37.79 to \textbf{41.84}, outperforming \texttt{CODEFILTER} by clear margins. These improvements align with our motivation: modeling evidence interactions helps retain complementary context while suppressing conflicting or redundant chunks.

Finally, the gains generalize beyond RepoEval. \textsc{RepoShapley} also delivers consistent improvements on long-context and chunk-level benchmarks, on CCLongEval Chunk it improves ES from 77.26 to \textbf{81.29} on SC-Base-7B and from 74.19 to \textbf{78.59} on Llama-13B, indicating that the learned keep and drop policy transfers across evaluation granularities and context regimes.

Although \textsc{RepoShapley} introduces additional computation during offline labeling, its inference-time overhead remains modest. As shown in Table~\ref{tab:ablation_component}, \textsc{RepoShapley} runs at 1053 ms/req, which is comparable to \texttt{CODEFILTER} (947 ms) and within the same runtime scale as RepoFormer (661 ms).

\noindent\textbf{Robustness to retriever choice.}
To verify that gains are not tied to sparse retrieval, we replace Jaccard with UniXcoder~\cite{guo2022unixcoder} as a dense retriever on SC-Base-7B (RepoEval-Line). As shown in Table~\ref{tab:dense_retrieval}, \textsc{RepoShapley} improves over Full-Retrieve under both retrievers, and the absolute gains are comparable (+8.80 ES with Jaccard vs. +7.69 ES with UniXcoder), confirming that coalition-aware filtering generalizes across retrieval paradigms. Appendix~\ref{app:sota_infilling} further shows that \textsc{RepoShapley} remains effective as an external selective-RAG policy for stronger modern code LMs, and Appendix~\ref{app:multilingual} validates cross-lingual generalization on Java, C\#, and TypeScript.

\begin{table}[t]
\centering
\small
\caption{Retriever comparison on SC-Base-7B (RepoEval-Line, Infilling).}
\label{tab:dense_retrieval}
\begin{tabular}{l|cc|cc}
\toprule
\multirow{2}{*}{\textbf{Method}} & \multicolumn{2}{c|}{\textbf{Jaccard}} & \multicolumn{2}{c}{\textbf{UniXcoder}} \\
\cmidrule(lr){2-3}\cmidrule(lr){4-5}
 & EM & ES & EM & ES \\
\midrule
Full-Retrieve & 58.26 & 77.79 & 62.71 & 80.45 \\
\textsc{RepoShapley} & \textbf{65.81} & \textbf{86.59} & \textbf{68.30} & \textbf{88.14} \\
\bottomrule
\end{tabular}
\end{table}

\subsection{Ablation Study \& Analysis}

We study how each component of \textsc{RepoShapley} affects performance by ablating (A) the offline labeling pipeline and (B) the online distillation strategy on StarCoderBase-1B with RepoEval (Table~\ref{tab:ablation_component}).

\noindent\textbf{Coalition-aware labeling matters.}
Ablations in Part~A show that modeling interactions is necessary for reliable filtering. Using Shapley signs without post-verification (Row~1) causes a large drop, indicating that signed marginal effects alone are not stable under prerequisite dependencies. Replacing coalition-based attribution with single-chunk probing (Row~2) also hurts performance, suggesting that independent scores miss synergy among chunks. Simplifying the surrogate utility by removing the sigmoid (Row~3) or using uniform weights (Row~4) further degrades results, supporting our design for capturing saturation and conflict effects.

\noindent\textbf{Distillation and triggering improve inference.}
Part~B shows that the training design is essential. Training with selection-only signals (Row~5) fails to produce usable code, while generation-only training (Row~6) lags behind the full model due to residual noise. Removing the trigger (Row~7) yields similar accuracy but increases latency, confirming that the learned trigger reduces unnecessary retrieval while maintaining generation quality.

\noindent\textbf{Sensitivity to surrogate scale $\beta$.}
We vary $\beta$ on RepoEval-Line (SC-Base-1B) and find that ES is robust for $\beta \in [0.5, 2.0]$, peaking at $\beta{=}1.0$ (82.78\%) and degrading at both extremes (78.5\% at $\beta{=}0.1$, 80.4\% at $\beta{=}5.0$). Small $\beta$ under-separates positive and negative chunks, while large $\beta$ saturates the sigmoid. We adopt $\beta{=}1.0$ as the default.
\begin{figure}[t]
    \centering
    \includegraphics[width=1\linewidth]{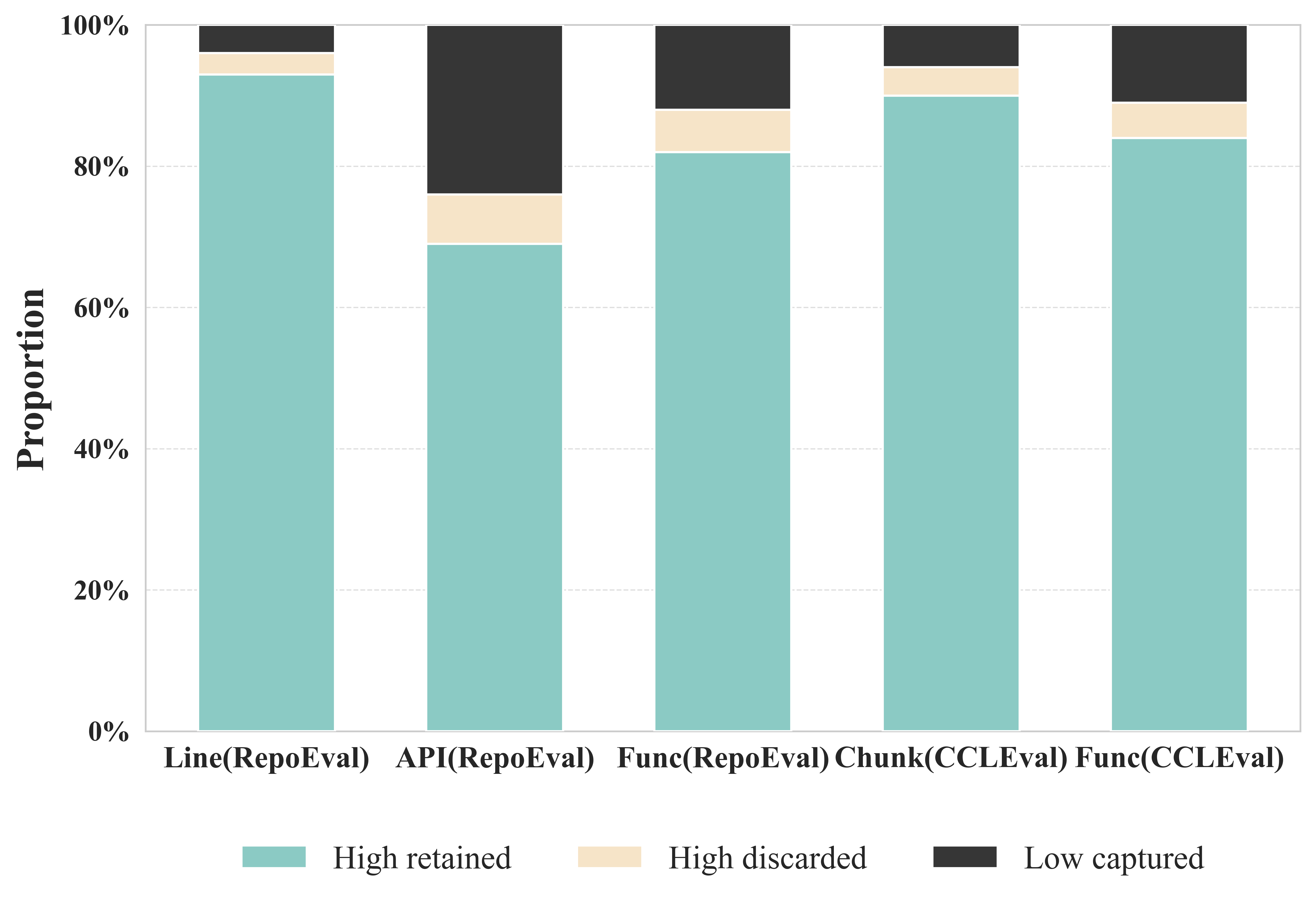}
    \caption{Breakdown of chunk selection decisions across benchmarks. \textbf{High retained}: consensus. \textbf{High discarded} and \textbf{Low captured}: corrections by RepoShapley.}
    \label{fig:abstention_analysis}
\end{figure}

\noindent\textbf{Selection behavior analysis.}
To understand what RepoShapley actually keeps and discards, we partition retrieved candidates by combining retriever score (high vs.\ low) with the keep/drop decision. As shown in Figure~\ref{fig:abstention_analysis}, on relatively local tasks (RepoEval Line, CCLEval Chunk), RepoShapley agrees with the retriever on most chunks (90\% high retained) while selectively correcting a small fraction. As task complexity increases, the corrections become more substantial: on RepoEval API, 7\% of high-score chunks are discarded and 24\% of kept chunks come from the low-score tail, confirming that interaction-aware filtering recovers individually under-ranked but coalition-critical evidence.

\begin{figure}[t]
    \centering
    \includegraphics[width=1\linewidth]{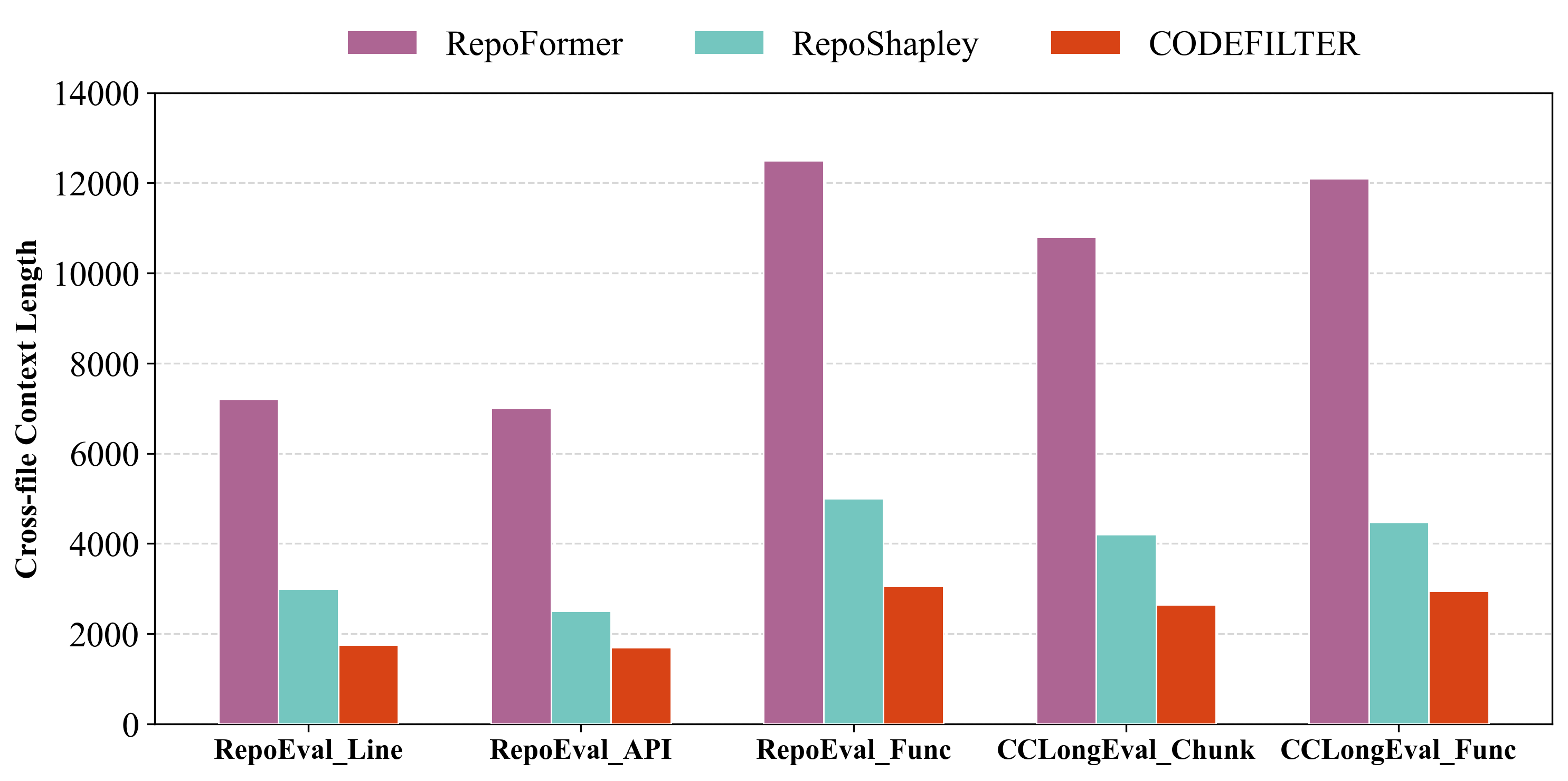}
    \caption{Retained cross-file context lengths. RepoFormer keeps the most tokens, \texttt{CODEFILTER} is most aggressive, and \textsc{RepoShapley} balances pruning with coverage.}
    \label{fig:context_length_dist}
\end{figure}

\noindent\textbf{Context length distribution.}
Figure~\ref{fig:context_length_dist} compares the number of cross-file tokens retained after filtering. RepoFormer retains the longest contexts (exceeding 12k tokens on function tasks), carrying substantial redundancy. \texttt{CODEFILTER} prunes most aggressively but may remove weak yet necessary dependencies. RepoShapley lies between the two, shortening contexts relative to RepoFormer while keeping complementary chunks that appear low-signal in isolation, achieving a favorable trade-off between token overhead and semantic coverage.

\section{Conclusion}
\label{sec:conclusion}

In this work, we study repository-level retrieval control under strong evidence interaction effects. We propose ChunkShapley, an offline labeling pipeline that estimates signed single-chunk effects, uses a structured surrogate game to capture saturation and conflict, and applies bounded post-verification to align selected coalitions with the generator's decoding behavior. The resulting Shapley-style labels are then distilled into \textsc{RepoShapley}, which performs retrieval triggering and chunk selection through discrete control tokens at inference time. Across benchmarks, backbones, retrievers, and prompting settings, \textsc{RepoShapley} consistently improves completion quality while reducing harmful or unnecessary context, showing that coalition-aware supervision is an effective foundation for repository-level code completion.

\section*{Limitations}
Our method has several limitations. First, the surrogate utility is built from single-chunk probes and may miss higher-order interactions where multiple individually weak chunks become useful only jointly; bounded post-verification mitigates but cannot fully guarantee recovery. Second, offline labeling enumerates $2^K$ subsets under the surrogate game ($K{=}10$), so the cost grows exponentially and limits scalability to larger retrieval budgets. Third, our framework uses a fixed chunking granularity, which affects the player set: finer granularity increases cost, while coarser granularity may mask intra-chunk interactions. Fourth, the verification stage is tied to greedy decoding with ES/EM; coalition choices may vary under different decoding strategies or objectives. Finally, the labeling pipeline requires multiple teacher-forced forward passes and decoding runs, increasing offline computation. Dataset licenses and code availability details are provided in Appendix~\ref{app:license}.

\section*{Acknowledgments}
This work is supported in part by the Guangdong Basic and Applied Basic Research Foundation under Grant No. 2025A1515012968, in part by the Shenzhen Science and Technology Program under Grant No. JCYJ20240813113502004, in part by the National Natural Science Foundation of China under Grant No. 62001412, in part by Shenzhen Stability Science Program 2023, in part by the Guangdong Provincial Key Laboratory of Future Networks of Intelligence (Grant No. 2022B1212010001), and in part by the Shenzhen Key Lab of Crowd Intelligence Empowered Low-Carbon Energy Network (Grant No. ZDSYS20220606100601002).

\bibliography{main}

@inproceedings{brown2020language,
 author = {Brown, Tom and Mann, Benjamin and Ryder, Nick and Subbiah, Melanie and Kaplan, Jared D and Dhariwal, Prafulla and Neelakantan, Arvind and Shyam, Pranav and Sastry, Girish and Askell, Amanda and Agarwal, Sandhini and Herbert-Voss, Ariel and Krueger, Gretchen and Henighan, Tom and Child, Rewon and Ramesh, Aditya and Ziegler, Daniel and Wu, Jeffrey and Winter, Clemens and Hesse, Chris and Chen, Mark and Sigler, Eric and Litwin, Mateusz and Gray, Scott and Chess, Benjamin and Clark, Jack and Berner, Christopher and McCandlish, Sam and Radford, Alec and Sutskever, Ilya and Amodei, Dario},
 booktitle = {Advances in Neural Information Processing Systems},
 editor = {H. Larochelle and M. Ranzato and R. Hadsell and M.F. Balcan and H. Lin},
 pages = {1877--1901},
 publisher = {Curran Associates, Inc.},
 title = {Language Models are Few-Shot Learners},
 url = {https://proceedings.neurips.cc/paper_files/paper/2020/file/1457c0d6bfcb4967418bfb8ac142f64a-Paper.pdf},
 volume = {33},
 year = {2020}
}

@inproceedings{wei2022chain,
 author = {Wei, Jason and Wang, Xuezhi and Schuurmans, Dale and Bosma, Maarten and ichter, brian and Xia, Fei and Chi, Ed and Le, Quoc V and Zhou, Denny},
 booktitle = {Advances in Neural Information Processing Systems},
 editor = {S. Koyejo and S. Mohamed and A. Agarwal and D. Belgrave and K. Cho and A. Oh},
 pages = {24824--24837},
 publisher = {Curran Associates, Inc.},
 title = {Chain-of-Thought Prompting Elicits Reasoning in Large Language Models},
 url = {https://proceedings.neurips.cc/paper_files/paper/2022/file/9d5609613524ecf4f15af0f7b31abca4-Paper-Conference.pdf},
 volume = {35},
 year = {2022}
}

@article{chen2021evaluating,
  author       = {Mark Chen and
                  Jerry Tworek and
                  Heewoo Jun and
                  Qiming Yuan and
                  Henrique Pond{\'{e}} de Oliveira Pinto and
                  Jared Kaplan and
                  Harri Edwards and
                  Yuri Burda and
                  Nicholas Joseph and
                  Greg Brockman and
                  Alex Ray and
                  Raul Puri and
                  Gretchen Krueger and
                  Michael Petrov and
                  Heidy Khlaaf and
                  Girish Sastry and
                  Pamela Mishkin and
                  Brooke Chan and
                  Scott Gray and
                  Nick Ryder and
                  Mikhail Pavlov and
                  Alethea Power and
                  Lukasz Kaiser and
                  Mohammad Bavarian and
                  Clemens Winter and
                  Philippe Tillet and
                  Felipe Petroski Such and
                  Dave Cummings and
                  Matthias Plappert and
                  Fotios Chantzis and
                  Elizabeth Barnes and
                  Ariel Herbert{-}Voss and
                  William Hebgen Guss and
                  Alex Nichol and
                  Alex Paino and
                  Nikolas Tezak and
                  Jie Tang and
                  Igor Babuschkin and
                  Suchir Balaji and
                  Shantanu Jain and
                  William Saunders and
                  Christopher Hesse and
                  Andrew N. Carr and
                  Jan Leike and
                  Joshua Achiam and
                  Vedant Misra and
                  Evan Morikawa and
                  Alec Radford and
                  Matthew Knight and
                  Miles Brundage and
                  Mira Murati and
                  Katie Mayer and
                  Peter Welinder and
                  Bob McGrew and
                  Dario Amodei and
                  Sam McCandlish and
                  Ilya Sutskever and
                  Wojciech Zaremba},
  title        = {Evaluating Large Language Models Trained on Code},
  journal      = {CoRR},
  volume       = {abs/2107.03374},
  year         = {2021},
  url          = {https://arxiv.org/abs/2107.03374},
  eprinttype   = {arXiv},
  eprint       = {2107.03374},
  bibsource    = {dblp computer science bibliography, https://dblp.org}
}

@inproceedings{
jimenez2024swe,
title={{SWE}-bench: Can Language Models Resolve Real-world Github Issues?},
author={Carlos E Jimenez and John Yang and Alexander Wettig and Shunyu Yao and Kexin Pei and Ofir Press and Karthik R Narasimhan},
booktitle={The Twelfth International Conference on Learning Representations},
year={2024},
url={https://openreview.net/forum?id=VTF8yNQM66}
}

@inproceedings{ding2024cocomic,
  title={Cocomic: Code completion by jointly modeling in-file and cross-file context},
  author={Ding, Yangruibo and Wang, Zijian and Ahmad, Wasi and Ramanathan, Murali Krishna and Nallapati, Ramesh and Bhatia, Parminder and Roth, Dan and Xiang, Bing},
  booktitle={Proceedings of the 2024 Joint International Conference on Computational Linguistics, Language Resources and Evaluation (LREC-COLING 2024)},
  pages={3433--3445},
  year={2024}
}

@article{liu2024lost,
  title={Lost in the middle: How language models use long contexts},
  author={Liu, Nelson F and Lin, Kevin and Hewitt, John and Paranjape, Ashwin and Bevilacqua, Michele and Petroni, Fabio and Liang, Percy},
  journal={Transactions of the Association for Computational Linguistics},
  volume={12},
  pages={157--173},
  year={2024}
}

@inproceedings{yoran2024making,
 author = {Yoran, Ori and Wolfson, Tomer and Ram, Ori and Berant, Jonathan },
 booktitle = {International Conference on Learning Representations},
 editor = {B. Kim and Y. Yue and S. Chaudhuri and K. Fragkiadaki and M. Khan and Y. Sun},
 pages = {29862--29883},
 title = {Making Retrieval-Augmented Language Models Robust to Irrelevant Context},
 url = {https://proceedings.iclr.cc/paper_files/paper/2024/file/8011b23e1dc3f57e1b6211ccad498919-Paper-Conference.pdf},
 volume = {2024},
 year = {2024}
}

@inproceedings{shi2023large,
  title={Large language models can be easily distracted by irrelevant context},
  author={Shi, Freda and Chen, Xinyun and Misra, Kanishka and Scales, Nathan and Dohan, David and Chi, Ed H and Sch{\"a}rli, Nathanael and Zhou, Denny},
  booktitle={International Conference on Machine Learning},
  pages={31210--31227},
  year={2023},
  organization={PMLR}
}

@inproceedings{
xu2024recomp,
title={{RECOMP}: Improving Retrieval-Augmented {LM}s with Context Compression and Selective Augmentation},
author={Fangyuan Xu and Weijia Shi and Eunsol Choi},
booktitle={The Twelfth International Conference on Learning Representations},
year={2024},
url={https://openreview.net/forum?id=mlJLVigNHp}
}

@inproceedings{bertsch2025context,
  title={In-context learning with long-context models: An in-depth exploration},
  author={Bertsch, Amanda and Ivgi, Maor and Xiao, Emily and Alon, Uri and Berant, Jonathan and Gormley, Matthew R and Neubig, Graham},
  booktitle={Proceedings of the 2025 Conference of the Nations of the Americas Chapter of the Association for Computational Linguistics: Human Language Technologies (Volume 1: Long Papers)},
  pages={12119--12149},
  year={2025}
}

@article{yan2024corrective,
  title={Corrective Retrieval Augmented Generation},
  author={Yan, Shi-Qi and Gu, Jia-Chen and Zhu, Yun and Ling, Zhen-Hua},
  journal={arXiv preprint arXiv:2401.15884},
  year={2024}
}

@inproceedings{
Khandelwal2020Generalization,
title={Generalization through Memorization: Nearest Neighbor Language Models},
author={Urvashi Khandelwal and Omer Levy and Dan Jurafsky and Luke Zettlemoyer and Mike Lewis},
booktitle={International Conference on Learning Representations},
year={2020},
url={https://openreview.net/forum?id=HklBjCEKvH}
}

@inproceedings{
wei2025instructrag,
title={Instruct{RAG}: Instructing Retrieval-Augmented Generation via Self-Synthesized Rationales},
author={Zhepei Wei and Wei-Lin Chen and Yu Meng},
booktitle={The Thirteenth International Conference on Learning Representations},
year={2025},
url={https://openreview.net/forum?id=P1qhkp8gQT}
}

@inproceedings{DingNIPS2024SemCoder,
 author = {Ding, Yangruibo and Peng, Jinjun and Min, Marcus J. and Kaiser, Gail and Yang, Junfeng and Ray, Baishakhi},
 booktitle = {Advances in Neural Information Processing Systems},
 doi = {10.52202/079017-1927},
 editor = {A. Globerson and L. Mackey and D. Belgrave and A. Fan and U. Paquet and J. Tomczak and C. Zhang},
 pages = {60275--60308},
 publisher = {Curran Associates, Inc.},
 title = {SemCoder: Training Code Language Models with Comprehensive Semantics Reasoning},
 url = {https://proceedings.neurips.cc/paper_files/paper/2024/file/6efcc7fd8efeee29a050a79c843c90e0-Paper-Conference.pdf},
 volume = {37},
 year = {2024}
}

@inproceedings{
kang2024crag,
title={C-{RAG}: Certified Generation Risks for Retrieval-Augmented Language Models},
author={Mintong Kang and Nezihe Merve G{\"u}rel and Ning Yu and Dawn Song and Bo Li},
booktitle={Forty-first International Conference on Machine Learning},
year={2024},
url={https://openreview.net/forum?id=FMa4c5NhOe}
}

@article{shrivastava2023repofusion,
  title={Repofusion: Training code models to understand your repository},
  author={Shrivastava, Disha and Kocetkov, Denis and De Vries, Harm and Bahdanau, Dzmitry and Scholak, Torsten},
  journal={arXiv preprint arXiv:2306.10998},
  year={2023}
}

@inproceedings{
bairi2023codeplan,
title={CodePlan: Repository-level Coding using {LLM}s and Planning},
author={Ramakrishna Bairi and Atharv Sonwane and Aditya Kanade and Vageesh D C and Arun Iyer and Suresh Parthasarathy and Sriram Rajamani and B. Ashok and Shashank Shet},
booktitle={NeurIPS 2023 Foundation Models for Decision Making Workshop},
year={2023},
url={https://openreview.net/forum?id=d0A2pc2kFp}
}

@article{gao2023retrieval,
  title={Retrieval-augmented generation for large language models: A survey},
  author={Gao, Yunfan and Xiong, Yun and Gao, Xinyu and Jia, Kangxiang and Pan, Jinliu and Bi, Yuxi and Dai, Yixin and Sun, Jiawei and Wang, Haofen and Wang, Haofen},
  journal={arXiv preprint arXiv:2312.10997},
  volume={2},
  number={1},
  year={2023}
}

@article{lewis2020RAG,
  title={Retrieval-augmented generation for knowledge-intensive nlp tasks},
  author={Lewis, Patrick and Perez, Ethan and Piktus, Aleksandra and Petroni, Fabio and Karpukhin, Vladimir and Goyal, Naman and K{\"u}ttler, Heinrich and Lewis, Mike and Yih, Wen-tau and Rockt{\"a}schel, Tim and others},
  journal={Advances in neural information processing systems},
  volume={33},
  pages={9459--9474},
  year={2020}
}

@inproceedings{izacard2021RAG,
  title={Leveraging passage retrieval with generative models for open domain question answering},
  author={Izacard, Gautier and Grave, Edouard},
  booktitle={Proceedings of the 16th conference of the european chapter of the association for computational linguistics: main volume},
  pages={874--880},
  year={2021}
}

@inproceedings{zhang2023repocoder,
  title={RepoCoder: Repository-Level Code Completion Through Iterative Retrieval and Generation},
  author={Zhang, Fengji and Chen, Bei and Zhang, Yue and Keung, Jacky and Liu, Jin and Zan, Daoguang and Mao, Yi and Lou, Jian-Guang and Chen, Weizhu},
  booktitle={The 2023 Conference on Empirical Methods in Natural Language Processing},
  year ={2023}
}

@article{ding2023crosscodeeval,
  title={Crosscodeeval: A diverse and multilingual benchmark for cross-file code completion},
  author={Ding, Yangruibo and Wang, Zijian and Ahmad, Wasi and Ding, Hantian and Tan, Ming and Jain, Nihal and Ramanathan, Murali Krishna and Nallapati, Ramesh and Bhatia, Parminder and Roth, Dan and others},
  journal={Advances in Neural Information Processing Systems},
  volume={36},
  pages={46701--46723},
  year={2023}
}

@inproceedings{liu2024repobench,
  title={RepoBench: Benchmarking Repository-Level Code Auto-Completion Systems},
  author={Liu, Tianyang and Xu, Canwen and McAuley, Julian},
  booktitle={The Twelfth International Conference on Learning Representations},
  year={2024}
}

@inproceedings{cheng2024dataflow,
  title={Dataflow-Guided Retrieval Augmentation for Repository-Level Code Completion},
  author={Cheng, Wei and Wu, Yuhan and Hu, Wei},
  booktitle={Proceedings of the 62nd Annual Meeting of the Association for Computational Linguistics (Volume 1: Long Papers)},
  pages={7957--7977},
  year={2024}
}

@inproceedings{liu2024graphcoder,
  title={Graphcoder: Enhancing repository-level code completion via coarse-to-fine retrieval based on code context graph},
  author={Liu, Wei and Yu, Ailun and Zan, Daoguang and Shen, Bo and Zhang, Wei and Zhao, Haiyan and Jin, Zhi and Wang, Qianxiang},
  booktitle={Proceedings of the 39th IEEE/ACM International Conference on Automated Software Engineering},
  pages={570--581},
  year={2024}
}

@inproceedings{parvez2021retrieval,
  title={Retrieval augmented code generation and summarization},
  author={Parvez, Md Rizwan and Ahmad, Wasi and Chakraborty, Saikat and Ray, Baishakhi and Chang, Kai-Wei},
  booktitle={Findings of the Association for Computational Linguistics: EMNLP 2021},
  pages={2719--2734},
  year={2021}
}

@inproceedings{guu2020retrieval,
  title={Retrieval augmented language model pre-training},
  author={Guu, Kelvin and Lee, Kenton and Tung, Zora and Pasupat, Panupong and Chang, Mingwei},
  booktitle={International conference on machine learning},
  pages={3929--3938},
  year={2020},
  organization={PMLR}
}

@inproceedings{zhang2025coderag,
    title = "{C}ode{RAG}: Finding Relevant and Necessary Knowledge for Retrieval-Augmented Repository-Level Code Completion",
    author = "Zhang, Sheng  and
      Ding, Yifan  and
      Lian, Shuquan  and
      Song, Shun  and
      Li, Hui",
    editor = "Christodoulopoulos, Christos  and
      Chakraborty, Tanmoy  and
      Rose, Carolyn  and
      Peng, Violet",
    booktitle = "Proceedings of the 2025 Conference on Empirical Methods in Natural Language Processing",
    month = nov,
    year = "2025",
    address = "Suzhou, China",
    publisher = "Association for Computational Linguistics",
    url = "https://aclanthology.org/2025.emnlp-main.1187/",
    doi = "10.18653/v1/2025.emnlp-main.1187",
    pages = "23278--23288",
    ISBN = "979-8-89176-332-6"
}

@inproceedings{generation2024coderag,
    title = "{C}ode{RAG}-Bench: Can Retrieval Augment Code Generation?",
    author = "Wang, Zora Zhiruo  and
      Asai, Akari  and
      Yu, Xinyan Velocity  and
      Xu, Frank F.  and
      Xie, Yiqing  and
      Neubig, Graham  and
      Fried, Daniel",
    editor = "Chiruzzo, Luis  and
      Ritter, Alan  and
      Wang, Lu",
    booktitle = "Findings of the Association for Computational Linguistics: NAACL 2025",
    month = apr,
    year = "2025",
    address = "Albuquerque, New Mexico",
    publisher = "Association for Computational Linguistics",
    url = "https://aclanthology.org/2025.findings-naacl.176/",
    doi = "10.18653/v1/2025.findings-naacl.176",
    pages = "3199--3214",
    ISBN = "979-8-89176-195-7"
}

@article{yang2025deep,
  title={A deep dive into retrieval-augmented generation for code completion: Experience on wechat},
  author={Yang, Zezhou and Peng, Ting and Gao, Cuiyun and Wang, Chaozheng and Huang, Hailiang and Deng, Yuetang},
  journal={arXiv preprint arXiv:2507.18515},
  year={2025}
}

@inproceedings{jiang2023active,
  title={Active retrieval augmented generation},
  author={Jiang, Zhengbao and Xu, Frank F and Gao, Luyu and Sun, Zhiqing and Liu, Qian and Dwivedi-Yu, Jane and Yang, Yiming and Callan, Jamie and Neubig, Graham},
  booktitle={Proceedings of the 2023 Conference on Empirical Methods in Natural Language Processing},
  pages={7969--7992},
  year={2023}
}

@inproceedings{mallen2023not,
  title={When not to trust language models: Investigating effectiveness of parametric and non-parametric memories},
  author={Mallen, Alex and Asai, Akari and Zhong, Victor and Das, Rajarshi and Khashabi, Daniel and Hajishirzi, Hannaneh},
  booktitle={Proceedings of the 61st Annual Meeting of the Association for Computational Linguistics (Volume 1: Long Papers)},
  pages={9802--9822},
  year={2023}
}

@inproceedings{asai2024self,
  title={Self-rag: Learning to retrieve, generate, and critique through self-reflection},
  author={Asai, Akari and Wu, Zeqiu and Wang, Yizhong and Sil, Avirup and Hajishirzi, Hannaneh},
  booktitle={The Twelfth International Conference on Learning Representations},
  year={2024}
}

@article{yao2024adaptive,
  title={Adaptive control of retrieval-augmented generation for large language models through reflective tags},
  author={Yao, Chengyuan and Fujita, Satoshi},
  journal={Electronics},
  volume={13},
  number={23},
  pages={4643},
  year={2024},
  publisher={MDPI}
}

@inproceedings{Wu2024repoformer,
author = {Wu, Di and Ahmad, Wasi Uddin and Zhang, Dejiao and Ramanathan, Murali Krishna and Ma, Xiaofei},
title = {REPOFORMER: selective retrieval for repository-level code completion},
year = {2024},
publisher = {JMLR.org},
booktitle = {Proceedings of the 41st International Conference on Machine Learning},
articleno = {2183},
numpages = {21},
location = {Vienna, Austria},
series = {ICML'24}
}

@inproceedings{
li2025codefilter,
title={Impact-driven Context Filtering For Cross-file Code Completion},
author={Yanzhou Li and Shangqing Liu and Kangjie Chen and Tianwei Zhang and Yang Liu},
booktitle={Second Conference on Language Modeling},
year={2025},
url={https://openreview.net/forum?id=0Y2zXLFBji}
}

@article{shapley1953value,
  title={A value for n-person games},
  author={Shapley, Lloyd S},
  journal={Contributions to the Theory of Games},
  volume={2},
  pages={307--317},
  year={1953},
  publisher={Princeton University Press Princeton}
}

@inproceedings{ghorbani2019data,
  title={Data shapley: Equitable valuation of data for machine learning},
  author={Ghorbani, Amirata and Zou, James},
  booktitle={International conference on machine learning},
  pages={2242--2251},
  year={2019},
  organization={PMLR}
}

@inproceedings{sundararajan2017axiomatic,
  title={Axiomatic attribution for deep networks},
  author={Sundararajan, Mukund and Taly, Ankur and Yan, Qiqi},
  booktitle={International conference on machine learning},
  pages={3319--3328},
  year={2017},
  organization={PMLR}
}

@article{ye2025fair,
  title={Fair Document Valuation in LLM Summaries via Shapley Values},
  author={Ye, Zikun and Yoganarasimhan, Hema},
  journal={arXiv preprint arXiv:2505.23842},
  year={2025}
}

@inproceedings{Lundberg2017shapley,
author = {Lundberg, Scott M. and Lee, Su-In},
title = {A unified approach to interpreting model predictions},
year = {2017},
isbn = {9781510860964},
publisher = {Curran Associates Inc.},
address = {Red Hook, NY, USA},
booktitle = {Proceedings of the 31st International Conference on Neural Information Processing Systems},
pages = {4768–4777},
numpages = {10},
location = {Long Beach, California, USA},
series = {NIPS'17}
}

@article{nematov2025source,
  title={Source Attribution in Retrieval-Augmented Generation},
  author={Nematov, Ikhtiyor and Kalai, Tarik and Kuzmenko, Elizaveta and Fugagnoli, Gabriele and Sacharidis, Dimitris and Hose, Katja and Sagi, Tomer},
  journal={arXiv preprint arXiv:2507.04480},
  year={2025}
}

@inproceedings{xiao2025tokenshapley,
  title={TokenShapley: Token Level Context Attribution with Shapley Value},
  author={Xiao, Yingtai and Zhu, Yuqing and Samyoun, Sirat and Zhang, Wanrong and Wang, Jiachen T and Du, Jian},
  booktitle={Findings of the Association for Computational Linguistics: ACL 2025},
  pages={3882--3894},
  year={2025}
}

@article{kocetkov2023stack,
title={The Stack: 3 {TB} of permissively licensed source code},
author={Denis Kocetkov and Raymond Li and Loubna Ben allal and Jia LI and Chenghao Mou and Yacine Jernite and Margaret Mitchell and Carlos Mu{\~n}oz Ferrandis and Sean Hughes and Thomas Wolf and Dzmitry Bahdanau and Leandro Von Werra and Harm de Vries},
journal={Transactions on Machine Learning Research},
issn={2835-8856},
year={2023},
url={https://openreview.net/forum?id=pxpbTdUEpD},
note={}
}

@article{li2023starcoder,
  title={StarCoder: May the Source be With You!},
  author={Li, R and Allal, LB and Zi, Y and Muennighoff, N and Kocetkov, D and Mou, C and Marone, M and Akiki, C and Li, J and Chim, J and others},
  journal={Transactions on machine learning research},
  year={2023},
  publisher={OpenReview}
}

@article{roziere2023codellama,
  title={Code llama: Open foundation models for code},
  author={Roziere, Baptiste and Gehring, Jonas and Gloeckle, Fabian and Sootla, Sten and Gat, Itai and Tan, Xiaoqing Ellen and Adi, Yossi and Liu, Jingyu and Sauvestre, Romain and Remez, Tal and others},
  journal={arXiv preprint arXiv:2308.12950},
  year={2023}
}

@article{jaccard1912distribution,
  title={The distribution of the flora in the alpine zone. 1},
  author={Jaccard, Paul},
  journal={New phytologist},
  volume={11},
  number={2},
  pages={37--50},
  year={1912},
  publisher={Wiley Online Library}
}

@misc{Yang2024Qwen25,
  title        = {{Qwen2.5-Max}: Exploring the Intelligence of Large-scale {MoE} Model},
  author       = {{Qwen Team}},
  year         = {2025},
  month        = jan,
  howpublished = {\url{https://qwenlm.github.io/blog/qwen2.5-max/}},
  note         = {Accessed: 2026-04-18}
}

@article{guo2025deepseek,
  title={DeepSeek-R1 incentivizes reasoning in LLMs through reinforcement learning},
  author={Guo, Daya and Yang, Dejian and Zhang, Haowei and Song, Junxiao and Wang, Peiyi and Zhu, Qihao and Xu, Runxin and Zhang, Ruoyu and Ma, Shirong and Bi, Xiao and others},
  journal={Nature},
  volume={645},
  number={8081},
  pages={633--638},
  year={2025},
  publisher={Nature Publishing Group UK London}
}

@article{zeng2025glm,
  title={Glm-4.5: Agentic, reasoning, and coding (arc) foundation models},
  author={Zeng, Aohan and Lv, Xin and Zheng, Qinkai and Hou, Zhenyu and Chen, Bin and Xie, Chengxing and Wang, Cunxiang and Yin, Da and Zeng, Hao and Zhang, Jiajie and others},
  journal={arXiv preprint arXiv:2508.06471},
  year={2025}
}

@misc{openai2025gpt5,
  title        = {Introducing {GPT-5}},
  author       = {{OpenAI}},
  year         = {2025},
  month        = aug,
  howpublished = {\url{https://openai.com/index/introducing-gpt-5/}},
  note         = {Accessed: 2026-04-18}
}

@techreport{anthropic2025claude4,
  title       = {System Card Addendum: {Claude Opus 4.1}},
  author      = {{Anthropic}},
  institution = {Anthropic},
  year        = {2025},
  month       = aug,
  url         = {https://assets.anthropic.com/m/4c024b86c698d3d4/original/Claude-4-1-System-Card.pdf},
  note        = {Accessed: 2026-04-18}
}

@inproceedings{guo2022unixcoder,
    title = "{U}ni{X}coder: Unified Cross-Modal Pre-training for Code Representation",
    author = "Guo, Daya  and
      Lu, Shuai  and
      Duan, Nan  and
      Wang, Yanlin  and
      Zhou, Ming  and
      Yin, Jian",
    editor = "Muresan, Smaranda  and
      Nakov, Preslav  and
      Villavicencio, Aline",
    booktitle = "Proceedings of the 60th Annual Meeting of the Association for Computational Linguistics (Volume 1: Long Papers)",
    month = may,
    year = "2022",
    address = "Dublin, Ireland",
    publisher = "Association for Computational Linguistics",
    url = "https://aclanthology.org/2022.acl-long.499/",
    doi = "10.18653/v1/2022.acl-long.499",
    pages = "7212--7225"
}

@misc{liu2026adaptivepromptstructurefactorization,
      title={Adaptive Prompt Structure Factorization: A Framework for Self-Discovering and Optimizing Compositional Prompt Programs}, 
      author={Haoyue Liu and Zhichao Wang and Yongxin Guo and Haoran Shou and Xiaoying Tang},
      year={2026},
      eprint={2604.06699},
      archivePrefix={arXiv},
      primaryClass={cs.CL},
      url={https://arxiv.org/abs/2604.06699}, 
}

\clearpage

\appendix

\begin{table*}[t]

\centering
\setlength{\tabcolsep}{4pt} 
\renewcommand{\arraystretch}{0.9}

\caption{Code completion performance in the Left-to-Right setting.}
\vspace{-5pt}

\resizebox{\textwidth}{!}{%
\begin{tabular}{l|l| cccccc |ccc |cc}
\toprule
\multirow{3}{*}{\textbf{Model}} & \multirow{3}{*}{\textbf{Strategy}} & \multicolumn{6}{c}{\textbf{RepoEval}} & \multicolumn{3}{c}{\textbf{CCLongEval}} & \multicolumn{2}{c}{\textbf{CCEval}} \\
\cmidrule(lr){3-8} \cmidrule(lr){9-11} \cmidrule(lr){12-13}
 & & \multicolumn{2}{c}{\textbf{Line}} & \multicolumn{2}{c}{\textbf{API}} & \multicolumn{2}{c}{\textbf{Function}} & \multicolumn{2}{c}{\textbf{Chunk}} & \textbf{Func} & \multicolumn{2}{c}{\textbf{Line}} \\
\cmidrule(lr){3-4} \cmidrule(lr){5-6} \cmidrule(lr){7-8} \cmidrule(lr){9-10} \cmidrule(lr){11-11} \cmidrule(lr){12-13}
 & & EM & ES & EM & ES & UT & ES & EM & ES & ES & EM & ES \\ \midrule

\multirow{5}{*}{\textbf{SC-Base-1B}} 
& No-Retrieve & 33.42 & 57.88 & 28.54 & 57.36 & 16.55 & 40.21 & 22.45 & 53.05 & 39.88 & 16.54 & 54.47 \\
& Full-Retrieve & 44.52 & 66.21 & 36.95 & 64.77 & 21.30 & 48.55 & 31.12 & 63.49 & 45.36 & 20.12 & 58.21 \\
& RepoFormer & 46.12 & 68.33 & 37.44 & 66.12 & 23.45 & 50.12 & 32.55 & 65.12 & 46.88 & 22.45 & 60.33 \\
& \texttt{CODEFILTER} & 48.88 & 70.15 & 39.85 & 69.11 & 24.12 & 51.55 & 34.15 & 66.88 & 48.22 & 24.88 & 62.55 \\
\rowcolor{gray!20}\cellcolor{white} & \textbf{\textsc{RepoShapley}} & \textbf{54.21}\g{5.33} & \textbf{76.45}\g{6.30} & \textbf{45.66}\g{5.81} & \textbf{74.88}\g{5.77} & \textbf{29.85}\g{5.73} & \textbf{57.22}\g{5.67} & \textbf{40.55}\g{6.40} & \textbf{72.44}\g{5.56} & \textbf{54.12}\g{5.90} & \textbf{30.12}\g{5.24} & \textbf{68.95}\g{6.40} \\ \midrule

\multirow{5}{*}{\textbf{SC-Base-3B}} 
& No-Retrieve & 35.82 & 60.12 & 29.55 & 58.45 & 20.05 & 38.95 & 25.44 & 53.45 & 44.82 & 18.22 & 57.51 \\
& Full-Retrieve & 50.45 & 70.88 & 40.66 & 67.89 & 26.12 & 48.66 & 36.15 & 59.88 & 45.75 & 23.45 & 62.12 \\
& RepoFormer & 50.11 & 71.95 & 41.02 & 69.88 & 27.55 & 50.12 & 36.75 & 61.95 & 47.12 & 25.66 & 63.88 \\
& \texttt{CODEFILTER} & 53.12 & 73.66 & 43.15 & 73.12 & 27.88 & 51.22 & 38.05 & 61.55 & 48.88 & 27.45 & 65.12 \\
\rowcolor{gray!20}\cellcolor{white} & \textbf{\textsc{RepoShapley}} & \textbf{59.45}\g{6.33} & \textbf{79.11}\g{5.45} & \textbf{48.88}\g{5.73} & \textbf{79.55}\g{6.43} & \textbf{33.45}\g{5.57} & \textbf{57.88}\g{6.66} & \textbf{44.22}\g{6.17} & \textbf{67.12}\g{5.57} & \textbf{54.66}\g{5.78} & \textbf{33.15}\g{5.70} & \textbf{70.44}\g{5.32} \\ \midrule

\multirow{5}{*}{\textbf{SC-Base-7B}} 
& No-Retrieve & 38.15 & 62.12 & 31.45 & 59.88 & 21.88 & 39.95 & 29.45 & 58.55 & 53.45 & 19.68 & 59.00 \\
& Full-Retrieve & 51.22 & 71.55 & 42.45 & 68.33 & 28.15 & 50.45 & 41.55 & 64.88 & 48.75 & 24.55 & 63.45 \\
& RepoFormer & 50.88 & 70.45 & 40.88 & 72.66 & 28.05 & 48.55 & 41.05 & 64.75 & 48.15 & 26.88 & 65.12 \\
& \texttt{CODEFILTER} & 53.95 & 74.22 & 44.55 & 72.15 & 29.05 & 52.33 & 42.12 & 66.88 & 57.88 & 28.55 & 67.55 \\
\rowcolor{gray!20}\cellcolor{white} & \textbf{\textsc{RepoShapley}} & \textbf{59.88}\g{5.93} & \textbf{80.55}\g{6.33} & \textbf{50.12}\g{5.57} & \textbf{78.45}\g{6.30} & \textbf{34.66}\g{5.61} & \textbf{58.12}\g{5.79} & \textbf{48.45}\g{6.33} & \textbf{72.15}\g{5.27} & \textbf{63.45}\g{5.57} & \textbf{34.12}\g{5.57} & \textbf{73.22}\g{5.67} \\ \midrule

\multirow{5}{*}{\textbf{Llama-7B}} 
& No-Retrieve & 39.55 & 64.12 & 30.88 & 60.22 & 22.45 & 42.55 & 30.12 & 58.12 & 45.45 & 20.88 & 60.12 \\
& Full-Retrieve & 52.45 & 70.88 & 43.15 & 68.75 & 26.88 & 50.12 & 41.22 & 63.88 & 52.66 & 25.44 & 64.55 \\
& RepoFormer & 51.12 & 71.45 & 40.88 & 70.88 & 28.66 & 50.05 & 39.45 & 62.95 & 50.45 & 27.12 & 65.88 \\
& \texttt{CODEFILTER} & 53.66 & 73.12 & 43.88 & 73.55 & 29.88 & 50.88 & 41.88 & 65.45 & 53.55 & 29.45 & 67.88 \\
\rowcolor{gray!20}\cellcolor{white} & \textbf{\textsc{RepoShapley}} & \textbf{59.12}\g{5.46} & \textbf{78.66}\g{5.54} & \textbf{49.55}\g{5.67} & \textbf{79.12}\g{5.57} & \textbf{35.12}\g{5.24} & \textbf{56.45}\g{5.57} & \textbf{47.22}\g{5.34} & \textbf{71.05}\g{5.60} & \textbf{59.12}\g{5.57} & \textbf{35.66}\g{6.21} & \textbf{73.45}\g{5.57} \\ \midrule

\multirow{5}{*}{\textbf{Llama-13B}} 
& No-Retrieve & 41.55 & 65.12 & 31.22 & 60.66 & 24.12 & 43.55 & 31.55 & 57.66 & 46.12 & 21.88 & 61.45 \\
& Full-Retrieve & 54.22 & 74.45 & 44.66 & 71.88 & 28.95 & 51.45 & 43.22 & 68.12 & 50.22 & 26.55 & 66.12 \\
& RepoFormer & 52.45 & 71.55 & 43.95 & 71.66 & 28.66 & 51.12 & 43.55 & 67.88 & 52.45 & 28.12 & 67.45 \\
& \texttt{CODEFILTER} & 55.33 & 75.12 & 45.12 & 74.95 & 30.12 & 52.45 & 44.22 & 67.95 & 57.12 & 30.88 & 69.55 \\
\rowcolor{gray!20}\cellcolor{white} & \textbf{\textsc{RepoShapley}} & \textbf{61.12}\g{5.79} & \textbf{81.45}\g{6.33} & \textbf{51.55}\g{6.43} & \textbf{80.66}\g{5.71} & \textbf{36.45}\g{6.33} & \textbf{58.22}\g{5.77} & \textbf{50.12}\g{5.90} & \textbf{73.45}\g{5.50} & \textbf{62.88}\g{5.76} & \textbf{36.95}\g{6.07} & \textbf{75.12}\g{5.57} \\ \bottomrule
\end{tabular}%
}
\label{tab:left_to_right_results}
\end{table*}

\section{Details of dataset construction}
\subsection{Metrics Formulation}
\label{metrics}
We evaluate code completion quality using Exact Match (EM), Edit Similarity (ES), and Unit Tests (UT). Let $\hat{Y}$ be the generated code and $Y$ be the ground truth:

\begin{align*}
\text{EM} &= \mathbf{1}(\hat{Y} = Y)\\
\text{ES} &= [1 - \frac{\mathcal{D}(\hat{Y}, Y)}{\max(|\hat{Y}|, |Y|)}] \times 100\%\\
\text{UT} &= \mathbf{1}(\text{Pass}(\hat{Y}))
\end{align*}

where $\mathbf{1}(\cdot)$ is the indicator function, $\mathcal{D}(\cdot)$ denotes the Levenshtein distance, and $\text{Pass}(\cdot)$ returns true if the code passes all unit tests.

\subsection{Data Collection and Preprocessing}
\label{sec:appendix_a}

\paragraph{File-level filtering.}
We begin with conservative file hygiene to reduce retrieval noise and stabilize likelihood-based labeling.
We keep only \texttt{.py} files and discard files with fewer than $10$ non-empty lines.
To remove minified/generated blobs that distort sparse retrieval, we drop files whose maximum line length exceeds $300$ characters or whose average line length exceeds $120$ characters (computed after trimming trailing whitespace).
We further filter out non-code payloads by requiring alphanumeric density $\ge 0.35$ (ratio of letters/digits over all characters).
Finally, we exclude vendored or generated directories by path keywords, including \texttt{vendor/}, \texttt{third\_party/}, \texttt{site-packages/}, \texttt{dist/}, \texttt{build/}, \texttt{.venv/}, and \texttt{migrations/}.
All statistics are computed on UTF-8 decoded text (with a permissive fallback that drops undecodable bytes).

\paragraph{Repository-level filtering.}
We retain repositories with sufficient structure for cross-file interactions by requiring at least $8$ remaining Python files and total non-empty LOC between $300$ and $50{,}000$ after file-level filtering.
To avoid duplicate-heavy projects where top-$K$ retrieval collapses to repeated copies, we estimate the near-duplicate file ratio using \textbf{SimHash}.
Specifically, for each repository we compute SimHash fingerprints over a normalized UTF-8 text representation of each file (whitespace-collapsed, with trailing whitespace removed), and perform pairwise checks on up to the first $200$ files (\texttt{max\_files\_for\_dup\_check}$=200$).
We mark two files as near-duplicates if their SimHash Hamming distance is at most $3$ (\texttt{simhash\_hamming\_threshold}$=3$).
Repositories with more than $30\%$ near-duplicate files are discarded (\texttt{max\_dup\_ratio}$=0.3$).

We also enforce syntactic integrity by parsing a sampled subset of files with Python \texttt{ast.parse}.
Concretely, we uniformly sample up to $20$ files per repository (\texttt{ast\_sample\_k}$=20$) from the remaining Python files after file-level filtering, and compute the parse success rate on this sample.
Repositories with parse success rate $<70\%$ are removed (\texttt{min\_ast\_parse\_rate}$=0.7$).
For reproducibility, both the duplicate-check subsampling (when applicable) and AST sampling use a fixed random seed of $13$ (\texttt{seed}$=13$).

\subsection{Data labeling.}
We chunk the cross-file pool and construct retrieval queries for each target span~$Y$.
During offline label construction, we use two candidate-retrieval variants in equal proportion: a context-only query from the in-file context and an oracle-assisted query that also includes $Y$.
The oracle-assisted query is used only to retrieve candidate chunks before ChunkShapley scoring; $Y$ is never included in the model input, inference-time retrieval query, or benchmark evaluation.
For each query, we retrieve the top-$K$ candidate chunks and distill coalition-aware decisions into
chunk-wise \tokkeep/\tokdrop\ labels and a retrieval-control token.
Specifically, given $X_{\text{in}}=(X_p,X_s)$ and retrieved candidates $X_{\text{cc}}=(cc_1,\dots,cc_K)$,
we run our offline \textbf{ChunkShapley} pipeline to obtain a verification-selected coalition $S^\star \subseteq \{1,\dots,K\}$.
We first compute a teacher-forced baseline log-likelihood $\ell(\emptyset)$ and probe each chunk in isolation,
$\Delta_i=\ell(\{i\})-\ell(\emptyset)$, yielding a signed vote $y_i=\mathrm{sign}(\Delta_i)$ and weight $\omega_i=|\Delta_i|$.
We then define a lightweight surrogate game
$v_{\text{sur}}(S)=\sigma(\beta\sum_{i\in S}\omega_i y_i)-\sigma(0)$, where $\sigma(\cdot)$ is the sigmoid function,
and compute exact surrogate Shapley values by enumerating all $2^K$ coalitions (tractable for small $K$ and performed offline).
Finally, we verify a bounded set of Shapley-proposed coalitions using the frozen generator under decoding-time constraints
and select $S^\star$ that maximizes completion quality (lexicographically by ES then EM).
We treat $S^\star$ as the teacher subset and assign labels:
$Q(cc_i)=\tokkeep$ if $i\in S^\star$ and $\tokdrop$ otherwise.

To supervise retrieval triggering, we assign the retrieval-control token
$r^\star\in\{\tokneed,\tokdone\}$ by comparing the completion quality with and without
cross-file evidence. If the verified coalition provides negligible gain over the in-file-only completion
($\mathrm{ES}(\hat{Y}_{S^\star},Y)-\mathrm{ES}(\hat{Y}_{\emptyset},Y)\le\epsilon$), we set
$r^\star=\tokdone$; otherwise $r^\star=\tokneed$.
To ensure that retained evidence is meaningful, we filter instances by requiring the verification-selected coalition to
achieve ES $\ge\tau_{\text{es}}$.
Alg.~\ref{alg:reposhapley_labeling} summarizes the cross-file labeling procedure.

\begin{table}[t]
\centering
\small
\setlength{\tabcolsep}{4pt}
\renewcommand{\arraystretch}{1.05}
\caption{Default labeling and inference settings used unless otherwise specified. ES thresholds are reported in ES points.}
\label{tab:labeling_defaults}
\begin{tabular}{lcc}
\toprule
\textbf{Setting} & \textbf{Symbol} & \textbf{Default} \\
\midrule
Retrieved chunks & $K$ & $10$ \\
Chunk/query window & $w$ & $512$ tokens \\
Chunk stride & $s$ & $256$ tokens \\
Shapley prefix count & $N_v$ & $10$ \\
Verification scope & $L$ & $3$ \\
Surrogate scale & $\beta$ & $1.0$ \\
Minimum verified ES & $\tau_{\text{es}}$ & $50$ \\
Retrieval-gain margin & $\epsilon$ & $0$ \\
Trigger threshold & $t_c$ & $0.5$ \\
Labeling query mix & -- & 50/50 ctx./oracle \\
Inference query & -- & ctx.-only \\
Train/validation split & -- & repo-level $95/5$ \\
\bottomrule
\end{tabular}
\end{table}

\begin{algorithm}[t]
\small
\SetAlgoLined
\DontPrintSemicolon
\caption{\textbf{ChunkShapley}: Surrogate Shapley Attribution with Bounded Verification}
\label{alg:chunkshapley}
\KwIn{
In-file context $X_{\text{in}}$; ground-truth completion $Y$; retrieved chunks $X_{\text{cc}}=(cc_1,\dots,cc_K)$;\\
Frozen generator $G_\theta$; surrogate scale $\beta$; verification params $(N_v, L)$.
}
\KwOut{Verification-selected coalition $S^\star \subseteq \{1,\dots,K\}$; decoded completion $\hat{Y}_{S^\star}$; surrogate Shapley scores $\{\phi_i\}_{i=1}^K$.}

\BlankLine
$\ell(\emptyset)\leftarrow \frac{1}{|Y|}\log p_\theta(Y \mid X_{\text{in}})$\;
\For{$i \leftarrow 1$ \KwTo $K$}{
    $\ell(\{i\}) \leftarrow \frac{1}{|Y|}\log p_\theta(Y \mid X_{\text{in}}, \{cc_i\})$\;
    $\Delta_i \leftarrow \ell(\{i\})-\ell(\emptyset)$\;
    $y_i \leftarrow \mathrm{sign}(\Delta_i)$;\quad $\omega_i \leftarrow |\Delta_i|$\;
}

\BlankLine
\ForEach{$S \subseteq \{1,\dots,K\}$}{
    $g(S) \leftarrow \sum_{j\in S}\omega_j y_j$\;
    $v_{\text{sur}}(S) \leftarrow \sigma(\beta\, g(S))-\sigma(0)$\;
}

\BlankLine
\For{$i \leftarrow 1$ \KwTo $K$}{
    $\phi_i \leftarrow 0$\;
    \ForEach{$S \subseteq \{1,\dots,K\}\setminus\{i\}$}{
        $w(S) \leftarrow \frac{|S|!\,(K-|S|-1)!}{K!}$\;
        $\phi_i \leftarrow \phi_i + w(S)\big(v_{\text{sur}}(S\cup\{i\})-v_{\text{sur}}(S)\big)$\;
    }
}

\BlankLine
$\pi_\phi \leftarrow \mathrm{argsort}(\{\phi_i\},\ \mathrm{desc})$;\quad
$\pi_\Delta \leftarrow \mathrm{argsort}(\{\Delta_i\},\ \mathrm{desc})$\;
$\mathcal{C} \leftarrow \mathrm{BuildPool}(\pi_\phi,\pi_\Delta; N_v, L)$

\ForEach{$S \in \mathcal{C}$}{
    $\hat{Y}_S \leftarrow \mathrm{Decode}\big(G_\theta \mid X_{\text{in}}, X_S\big)$\;
    Compute $\mathrm{ES}(\hat{Y}_S,Y)$ and $\mathrm{EM}(\hat{Y}_S,Y)$\;
}
$S^\star \leftarrow \arg\max_{S\in\mathcal{C}}\big(\mathrm{ES}(\hat{Y}_S,Y),\ \mathrm{EM}(\hat{Y}_S,Y)\big)$

\Return $S^\star,\ \hat{Y}_{S^\star},\ \{\phi_i\}_{i=1}^K$\;
\end{algorithm}

\subsection{Labeling Algorithm Details}
\label{app:labeling_details}

Algorithm \ref{alg:reposhapley_labeling} outlines the process of deriving supervision signals from raw code repositories. First, top-$K$ candidate chunks $X_{\text{cc}}$ are retrieved based on the query window $Q$. We then utilize \textsc{ChunkShapley} to identify the optimal chunk subset $S^\star$ that maximizes generation quality relative to the ground truth $Y$.

The labeling logic follows three specific criteria:
\begin{enumerate}
    \item \textbf{Quality Control:} Instances are discarded if the optimal subset's performance falls below a minimum threshold $\tau_{\text{es}}$, ensuring training data quality.
    \item \textbf{Retrieval Label ($r^\star$):} We measure the performance gain of using external context ($S^\star$) versus the closed-book baseline ($\emptyset$). If the gain is negligible ($\le \epsilon$), the retrieval label is set to $\tokdone$; otherwise, it is $\tokneed$.
    \item \textbf{Selection Label ($q_i^\star$):} Individual chunks are labeled as $\tokkeep$ if they belong to the optimal subset $S^\star$, and $\tokdrop$ otherwise.
\end{enumerate}

\begin{algorithm}[t]
\small
\SetAlgoLined
\DontPrintSemicolon
\caption{\textsc{RepoShapley} Cross-file Labeling via ChunkShapley}
\label{alg:reposhapley_labeling}
\KwIn{
Repository cross-file pool $X_{\text{out}}$; in-file context $X_{\text{in}}=(X_p,X_s)$; target span $Y$;\\
Retriever $R$; frozen generator $G$; chunk window $w$; stride $s$; retrieve budget $K$;\\
verification params $(N_v,L)$ (as in Alg.~\ref{alg:chunkshapley}); thresholds $\tau_{\text{es}}$ and $\epsilon$;\\
query mode $m\in\{\mathrm{ctx},\mathrm{oracle}\}$.
}
\KwOut{
Labeled instance: retrieval label $r^\star\in\{\tokneed,\tokdone\}$ and
selection labels $(q_1^\star,\dots,q_K^\star)$ with $q_i^\star\in\{\tokkeep,\tokdrop\}$
}

\BlankLine
$Q_{\mathrm{ctx}} \leftarrow \mathrm{ContextWindow}(X_{\text{in}};\,w)$\;
\eIf{$m=\mathrm{oracle}$}{
$Q \leftarrow Q_{\mathrm{ctx}}\oplus Y$ \tcp*{labeling only}
}{
$Q \leftarrow Q_{\mathrm{ctx}}$ \tcp*{inference matched}
}
$\widetilde{X}_{\text{out}} \leftarrow \mathrm{chunkize}(X_{\text{out}};\, w,s)$\;
$X_{\text{cc}} \leftarrow R(Q,\widetilde{X}_{\text{out}})[1{:}K]$ \\
$(S^\star,\hat{Y}_{S^\star},\{\phi_i\}_{i=1}^K) \leftarrow \textbf{ChunkShapley}(X_{\text{in}},Y,X_{\text{cc}},G;\,N_v,L)$\;
$\hat{Y}_{\emptyset} \leftarrow G(X_{\text{in}})$ \\

\If{$\mathrm{ES}(\hat{Y}_{S^\star},Y) < \tau_{\text{es}}$}{
\Return \textbf{discard instance}\;
}

\eIf{$\mathrm{ES}(\hat{Y}_{S^\star},Y)-\mathrm{ES}(\hat{Y}_{\emptyset},Y)\le \epsilon$}{
$r^\star \leftarrow \tokdone$\;
}{
$r^\star \leftarrow \tokneed$\;
}

\For{$i\leftarrow 1$ \KwTo $K$}{
\eIf{$i\in S^\star$}{
$q_i^\star \leftarrow \tokkeep$\;
}{
$q_i^\star \leftarrow \tokdrop$\;
}
}

\BlankLine
\Return $r^\star,(q_1^\star,\dots,q_K^\star)$\;
\end{algorithm}

\section{Hyperparameter Optimization}
\label{app:hparam}

We tune training hyperparameters using \textbf{StarCoderBase-1B} as a proxy model to reduce search cost.
Unless otherwise specified, all other settings follow the main experimental setup (e.g., data split, prompt formats, max sequence length, and batching).

\paragraph{Search space.}
We conduct a grid search on the following space:
learning rate $\in \{1\times 10^{-5},\,2\times 10^{-5},\,5\times 10^{-5}\}$,
loss weight $\lambda \in \{0.2,\,1.0,\,2.0,\,5.0\}$,
training epochs $\in \{1,\,2,\,5\}$,
and warmup steps $\in \{50,\,100\}$.
Here $\lambda$ is applied to the retrieval-control and selection losses, i.e.,
$\lambda_R=\lambda_S=\lambda$ (while the generation loss uses unit weight).

\paragraph{Selection criterion.}
For each configuration, we evaluate code completion performance on the validation split using the same metrics as in the main experiments.
We select the best hyperparameters by maximizing the validation completion quality (with ES as the primary criterion and EM as a tie-breaker).

\paragraph{Final configuration.}
The selected hyperparameters are:
learning rate $2\times 10^{-5}$, $\lambda_R=\lambda_S=2.0$, epochs $=2$, warmup steps $=50$.
We reuse this configuration for all backbones in our experiments for consistency.

\begin{table}[t]
\centering
\small
\setlength{\tabcolsep}{6pt}
\renewcommand{\arraystretch}{1.1}
\caption{Hyperparameter search space and selected values (tuned on StarCoderBase-1B).}
\label{tab:hparam_search}
\begin{tabular}{lcc}
\toprule
\textbf{Hyperparameter} & \textbf{Search space} & \textbf{Selected} \\
\midrule
Learning rate & $\{1\mathrm{e}{-5},\,2\mathrm{e}{-5},\,5\mathrm{e}{-5}\}$ & $2\mathrm{e}{-5}$ \\
$\lambda$ ($\lambda_R=\lambda_S$) & $\{0.2,\,1.0,\,2.0,\,5.0\}$ & $2.0$ \\
Epochs & $\{1,\,2,\,5\}$ & $2$ \\
Warmup steps & $\{50,\,100\}$ & $50$ \\
\bottomrule
\end{tabular}
\end{table}

\section{Detailed Experiments}

\subsection[Ablation on Verification Scope L]{Ablation on Verification Scope ($L$)}
\label{app:ablation_top_l_text}

Table~\ref{tab:ablation_top_l} shows that expanding the verification scope from $L=0$ to $L=3$ brings the largest gains: moving beyond prefix-only verification substantially improves both EM and ES, and performance increases steadily up to the default $L=3$. In contrast, further enlarging the scope ($L>3$) yields only marginal improvements, despite a rapidly growing candidate pool and offline labeling cost. Therefore, we adopt $L=3$ as a practical default that captures most of the benefit of combinatorial probing.

\begin{table*}[t]
\centering
\caption{\textbf{Ablation on Verification Scope ($L$).} Impact of the Top-$L$ range used for combinatorial probing in the post-verification stage. We report the average size of the candidate pool $|\mathcal{C}|$, offline labeling latency per instance, and performance on SC-Base-1B.}
\label{tab:ablation_top_l}
\setlength{\tabcolsep}{6pt}
\renewcommand{\arraystretch}{1.1}
\begin{tabular}{c|cc|cc}
\toprule
\multirow{2}{*}{\textbf{Top-$L$}} & \multicolumn{2}{c|}{\textbf{Labeling Cost}} & \multicolumn{2}{c}{\textbf{Performance}} \\
\cmidrule(lr){2-3} \cmidrule(lr){4-5}
 & \textbf{Avg. Pool Size} ($|\mathcal{C}|$) & \textbf{Train Time per Sample} (s) & \textbf{EM} (\%) & \textbf{ES} (\%) \\
\midrule
0 (Prefix Only) & 8.41 & 33 & 45.15 & 70.86 \\ 
1  & 10.78 & 94 & 57.73 & 76.27 \\ 
2  & 12.56 & 187 & 59.15 & 78.40 \\ 
\rowcolor{gray!15} \textbf{3}(Default)  & 18.29 & 348 & 61.34 & 82.78 \\

4 & 25.07 & 671 & 61.70 & 83.22 \\
5 & 35.42 & 869 & 61.94 & 83.24 \\
7 & 68.94 & 1528 & 62.13 & 83.59 \\
10 (All) & 225 & 6823 & - & - \\
\bottomrule
\end{tabular}
\end{table*}

\subsection{Oracle Analysis}
\label{sec:appendix_oracle}

To validate the theoretical superiority of Shapley-based valuation over independent likelihood probing (as used in \texttt{CODEFILTER}), we conducted an Oracle study. We calculated the best possible Edit Similarity (ES) achievable if the model perfectly selected chunks according to the respective valuation methods (selecting top-$K$ chunks with score $>0$).

\begin{table}[t]
\centering
\small
\caption{\textbf{Oracle Performance Comparison.} We report the mean Best ES score achievable by selecting contexts based on oracle labels. \textsc{RepoShapley} (Oracle) demonstrates a significantly higher theoretical upper bound.}
\begin{tabular}{l|c}
\toprule
\textbf{Method} & \textbf{Oracle Best ES (\%)}  \\
\midrule
Full-Retrieve & 71.52  \\
\midrule
\texttt{CODEFILTER} (Oracle) & 85.23 \\
\textsc{RepoShapley (Oracle)} & \textbf{95.68} \\
\bottomrule
\end{tabular}
\label{tab:oracle_comparison}
\end{table}

As shown in Table \ref{tab:oracle_comparison}, the Shapley-based oracle outperforms the \texttt{CODEFILTER} oracle by \textbf{10.45} percentage points. This confirms that modelling chunk interactions, like synergy and conflict, is critical for repository-level code completion, as independent probing fails to identify chunks that are only useful when combined, for instance interface definitions and implementations.

\subsection[Sensitivity Analysis on Retrieval Budget K]{Sensitivity Analysis on Retrieval Budget $K$}
Table~\ref{tab:k_ablation} investigates the trade-off between completion performance and inference latency by varying the retrieval budget $K$ (i.e., the number of candidate chunks processed by ChunkShapley) on SC-Base-1B.
Increasing $K$ expands the search space for complementary evidence, potentially capturing more synergistic interactions. However, since our method involves exact Shapley estimation via subset enumeration, the computational cost grows exponentially with $K$.
Specifically, a small budget ($K=7$) yields low latency but fails to retrieve sufficient complementary pairs, resulting in suboptimal accuracy (52.16\% EM).
Conversely, increasing $K$ beyond 10 yields \textit{diminishing returns}; for instance, expanding to $K=13$ marginally improves ES by 0.51\% but causes latency to explode to over 3.5 seconds due to the combinatorial complexity of the surrogate game ($2^{13}$ subsets), rendering it impractical.
Crucially, $K=10$ achieves the optimal balance; we adopt it as the default setting to balance interaction coverage with inference efficiency.

\begin{table}[t]
\centering
\renewcommand{\arraystretch}{1.1}

\caption{Sensitivity analysis of the retrieval budget ($K$) on SC-Base-1B in the Infilling setting. We report Line Completion accuracy (EM/ES) and average inference latency. $K=10$ achieves the best trade-off between context coverage and computational cost.}
\vspace{-5pt}

\resizebox{\linewidth}{!}{%
\begin{tabular}{c|cc|c}
\toprule
\multirow{2}{*}{\textbf{Retrieval Size}} & \multicolumn{2}{c|}{\textbf{RepoEval-Line}} & {\textbf{Efficiency}} \\
\cmidrule(lr){2-3}\cmidrule(lr){4-4}
 & \textbf{EM } & \textbf{ES} & \textbf{Latency (ms)} \\ \midrule

7 & 52.16 & 70.44 & 513  \\
9 & 58.33 & 75.12 & 833  \\
\rowcolor{gray!20} \textbf{10 (Ours)} & \textbf{61.34} & \textbf{82.78} & \textbf{1053} \\
11 & 61.37 & 82.40 & 1924  \\
13 & 61.88 & 81.62 & 3539 \\ 
20 & 61.76 & 79.92 & 18825\\
\bottomrule

\end{tabular}%
}
\label{tab:k_ablation}
\end{table}

\subsection{Ablation Study on Coalition Utility Functions}
\label{app:ablation_utility}

In \textbf{ChunkShapley}, the choice of the characteristic function $v(S)$ is critical, as it defines the "value" distributed among retrieved chunks. We hypothesize that while task-specific metrics (like Exact Match) align perfectly with the final objective, they provide sparse and noisy signals for attribution. To validate the effectiveness of our Log-likelihood-based utility, we conduct an ablation study comparing it against task-metric-based utilities.

\paragraph{Experimental Setup.}
We compare three definitions of coalition utility $v(S)$:

\textbf{Log-likelihood Utility (Ours):}
    We use the normalized token-level log-probability gain under teacher forcing. This provides a continuous, dense signal reflecting the model's confidence:
    \begin{align*}
        v_{\text{log}}(S)
= \frac{1}{|Y|}\sum_{t=1}^{|Y|}
\Big(
\log p_\theta(y_t \mid y_{<t}, X_{\text{in}}, X_S) \\
-
\log p_\theta(y_t \mid y_{<t}, X_{\text{in}}, \emptyset)
\Big).
    \end{align*}

    \textbf{Exact Match (EM) Utility:}
    We define utility as the binary gain in obtaining a perfect prediction. This signal is discrete ($\in \{-1, 0, 1\}$) and highly sparse:
    \[ 
    \begin{aligned}
    v_{\text{EM}}(S) &= \mathbf{1}[\text{EM}(\hat{Y}_S, Y)=1]\\
    &\quad - \mathbf{1}[\text{EM}(\hat{Y}_{\emptyset}, Y)=1].
    \end{aligned}
    \]
    where $\hat{Y}_S$ is the greedy decoding result given context $S$.

    \textbf{Edit Similarity (ES) Utility:}
    We define utility based on the improvement in surface-level similarity. While continuous (0-100), ES is derived from discrete decoding steps and is non-differentiable:
    \begin{equation}
    v_{\text{ES}}(S) = \text{ES}(\hat{Y}_S, Y) - \text{ES}(\hat{Y}_{\emptyset}, Y)
    \end{equation}

\begin{table*}[t]
\centering
\setlength{\tabcolsep}{6pt}
\renewcommand{\arraystretch}{1.1}
\caption{Multilingual evaluation on CrossCodeEval (Line completion, ES / ID F1).}
\label{tab:multilingual}
\resizebox{\textwidth}{!}{%
\begin{tabular}{l|l|cc|cc|cc|cc}
\toprule
\multirow{2}{*}{\textbf{Model}} & \multirow{2}{*}{\textbf{Strategy}} & \multicolumn{2}{c|}{\textbf{Python}} & \multicolumn{2}{c|}{\textbf{Java}} & \multicolumn{2}{c|}{\textbf{C\#}} & \multicolumn{2}{c}{\textbf{TypeScript}} \\
\cmidrule(lr){3-4}\cmidrule(lr){5-6}\cmidrule(lr){7-8}\cmidrule(lr){9-10}
 & & ES & F1 & ES & F1 & ES & F1 & ES & F1 \\
\midrule
\multirow{2}{*}{SC-Base-7B} & Full-Retrieve & 52.74 & 43.42 & 58.78 & 49.53 & 63.72 & 55.64 & 51.28 & 42.29 \\
 & \textbf{RepoShapley-M} & \textbf{62.91} & \textbf{53.30} & \textbf{66.20} & \textbf{58.29} & \textbf{71.83} & \textbf{64.23} & \textbf{60.23} & \textbf{51.63} \\
\midrule
\multirow{2}{*}{Llama-13B} & Full-Retrieve & 50.20 & 42.27 & 60.23 & 50.44 & 62.80 & 55.86 & 48.06 & 41.24 \\
 & \textbf{RepoShapley-M} & \textbf{59.37} & \textbf{51.39} & \textbf{67.11} & \textbf{57.17} & \textbf{70.95} & \textbf{63.20} & \textbf{56.98} & \textbf{50.80} \\
\bottomrule
\end{tabular}%
}
\end{table*}

For all variants, we compute the exact Shapley values using the respective $v(S)$, select the optimal subset $S^\star$ using the verification strategy, and train the corresponding \textsc{RepoShapley} model.

\paragraph{Results and Analysis.} As shown in Table~\ref{tab:ablation_utility}, using log-likelihood as the utility function yields the best performance. We attribute the inferiority of metric-based utilities ($v_{\text{EM}}, v_{\text{ES}}$) to the \textbf{sparsity and high variance} of the signal. In code completion, a chunk might significantly improve the model's understanding (providing the correct variable type) without immediately flipping the final prediction to an exact match. $v_{\text{log}}$ captures this "partial credit," whereas $v_{\text{EM}}$ assigns zero value, leading to false negatives in attribution. Furthermore, generation-based metrics are sensitive to decoding dynamics, where a small change in context might drastically alter the greedy path, causing $v_{\text{metric}}(S)$ to fluctuate wildly. In contrast, teacher-forced log-probabilities provide a smoother and more robust estimation of marginal contribution.

\begin{table}[t]
    \centering
    \caption{\textbf{Ablation study on Utility Functions.} We report the performance of \textbf{\textsc{RepoShapley}} when trained with Shapley labels derived from different utility definitions. \emph{Log-likelihood} (Ours) significantly outperforms metric-based utilities due to the density and stability of the teacher-forcing signal.}
    \label{tab:ablation_utility}
    \resizebox{\linewidth}{!}{
    \begin{tabular}{l|c|cc}
    \toprule
    \textbf{Utility Function} & \textbf{Signal Type} & \textbf{EM} & \textbf{ES} \\
    \midrule
    \textbf{w/ $v_{\text{EM}}$} & Binary & 58.12 & 77.45 \\
    \textbf{w/ $v_{\text{ES}}$} & Discrete-Step & 59.03 & 78.10 \\
    \midrule
    \textbf{w/ $v_{\text{log}}$ (Ours)} & \textbf{Continuous} & \textbf{60.50} & \textbf{79.07} \\
    \bottomrule
    \end{tabular}
    }
\end{table}

\subsection{Filtering effectiveness via selective drop and counterfactual inverse.}
Following \texttt{CODEFILTER}~\cite{li2025codefilter}, we study whether attribution signals can reliably separate helpful from harmful retrieved context on RepoEval under the same Jaccard-based retriever. Table~\ref{tab:context_strategies_drop} reports two complementary interventions: \textbf{Selective (Drop)}, which removes chunks labeled as \tokdrop\ and keeps the remaining context, and \textbf{Inverse (Keep)}, which retains only the dropped chunks as a counterfactual diagnostic.

\begin{figure}[t]
            \centering
            \includegraphics[width=0.99\linewidth]{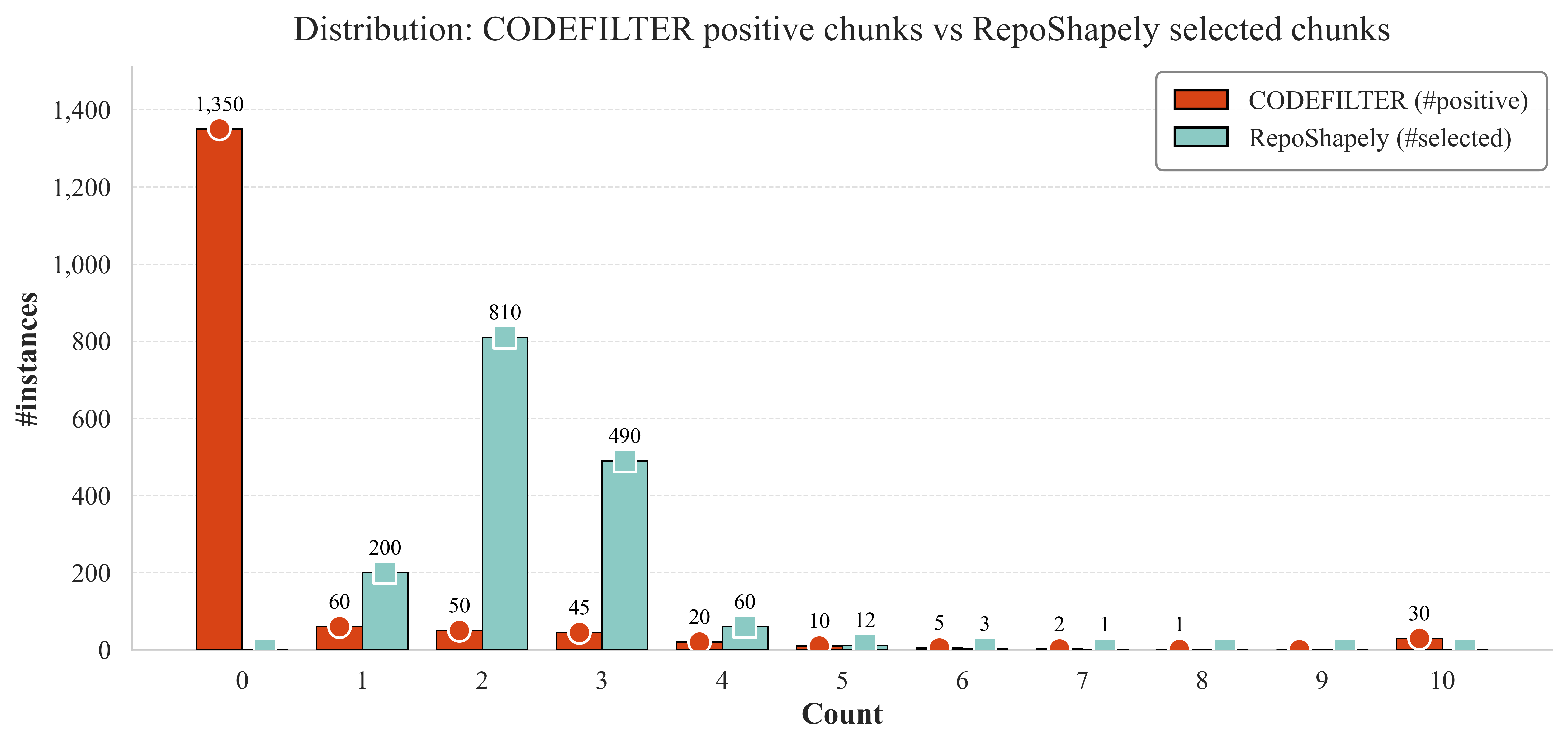}
            \caption{Distribution: \texttt{CODEFILTER} positive chunks vs. \textsc{RepoShapley} selected chunks}
            \label{fig:distribution}
\end{figure}

\textsc{RepoShapley} gains markedly from selective filtering, improving both EM and ES compared to the \texttt{CODEFILTER} counterpart. In contrast, the inverse setting substantially degrades performance for both methods, confirming that the dropped chunks are predominantly low-utility (e.g., redundant or misleading) rather than accidentally filtered-out evidence. Notably, Figure~\ref{fig:distribution} shows that \texttt{CODEFILTER}’s decisions are prone to brittle single-chunk thresholds: when individual signals are weak, it tends to label very few chunks as positive, effectively collapsing the available context. \textsc{RepoShapley} instead maintains a stable selection set, consistent with interaction-aware supervision that removes toxic context while preserving the evidence required for accurate repository-level completion.

\begin{table}[t]
\centering
\renewcommand{\arraystretch}{1} 
\caption[Impact of the filtering policy based on different attribution signals. Selective denotes removing chunks labeled as \texttt{<DROP>}, while Inverse retains only those chunks (to verify the toxicity of dropped content).]{Impact of the filtering policy based on different attribution signals. \textbf{Selective} denotes removing chunks labeled as \tokdrop, while \textbf{Inverse} retains only those chunks (to verify the toxicity of dropped content).}
\label{tab:context_strategies_drop}

\setlength{\tabcolsep}{3pt}
\begin{tabular}{l|cc|>{\columncolor{gray!20}}c>{\columncolor{gray!20}}c}
\toprule
\multirow{2}{*}{\textbf{Strategy}} &
\multicolumn{2}{c}{\texttt{CODEFILTER}} &
\multicolumn{2}{c}{\textsc{RepoShapley}} \\
\cmidrule(lr){2-3}\cmidrule(lr){4-5}
 & \textbf{EM} & \textbf{ES} & \textbf{EM} & \textbf{ES} \\
\midrule


\textbf{Selective} (Drop) & 49.31 & 57.50 & \textbf{54.79}$\uparrow$ & \textbf{78.62}$\uparrow$ \\ \midrule

Inverse (Keep) & 33.17 & 40.26 & 34.72 & 41.15 \\ 

\bottomrule
\end{tabular}
\end{table}

\begin{table}[t]
\renewcommand{\arraystretch}{1} 
    \centering
    \caption{\textbf{Impact of Proposal Mechanism.} Comparison between using Single-Chunk $\Delta$ vs. Coalition Shapley $\phi$ to generate candidates for the verification step.}
    \label{tab:delta_ablation}
    \resizebox{\linewidth}{!}{
    \begin{tabular}{l|c|c}
    \toprule
    \textbf{Proposal Method} & \textbf{Metric} & \textbf{Verified ES (\%)} \\
    \midrule
    Delta ($\Delta$) + Verify & ES  & 74.21 \\
    \textbf{\textsc{RepoShapley} ($\phi$) + Verify} & \textbf{ES} & \textbf{82.78} \\
    \bottomrule
    \end{tabular}
    }
\end{table}
\vspace{-8pt}

\begin{table*}[t]
\centering
\setlength{\tabcolsep}{6pt}
\setlength{\tabcolsep}{12pt} 
\renewcommand{\arraystretch}{1.1} 

\caption{Accuracy of modern code LMs as the generation model and with \textsc{RepoShapley} as the policy model for selective RAG in Infilling setting. We compare DeepSeek \cite{guo2025deepseek}, Qwen2.5-Max \cite{Yang2024Qwen25}, GLM \cite{zeng2025glm}, GPT-5 \cite{openai2025gpt5}, and Claude Opus 4.1 \cite{anthropic2025claude4}. \textbf{For closed-source models, we use the official APIs for generation.}}

\vspace{-5pt}

\begin{tabular}{l|l|cc|cc}
\toprule
\multirow{2}{*}{\textbf{Model}} & \multirow{2}{*}{\textbf{Strategy}} &\multicolumn{2}{c}{\textbf{RepoEval-Line}} & \multicolumn{2}{c}{\textbf{RepoEval-API}} \\
\cmidrule(l){3-6}
 & & \textbf{EM} & \textbf{ES} & \textbf{EM} & \textbf{ES} \\ \midrule

\multirow{2}{*}{StarCoderBase-7B} 
& Full-Retrieve & 58.26 & 77.79 & 50.38 & 75.01 \\
& \textbf{RepoShapley} & \textbf{65.81} & \textbf{86.59} & \textbf{58.79} & \textbf{84.11} \\ \midrule

\multirow{2}{*}{CodeLlama-13B} 
& Full-Retrieve & 61.41 & 79.29 & 49.81 & 77.41 \\
& \textbf{RepoShapley} & \textbf{68.89} & \textbf{87.11} & \textbf{57.66} & \textbf{83.41} \\ \midrule

\multirow{2}{*}{DeepSeek-R1} 
& Full-Retrieve & 62.94 & 80.09 & 51.61 & 80.93 \\
& \textbf{RepoShapley} & \textbf{68.79} & \textbf{87.92} & \textbf{58.96} & \textbf{84.07} \\ \midrule

\multirow{2}{*}{Qwen2.5-Max} 
& Full-Retrieve & 63.38 & 81.91 & 54.72 & 81.39 \\
& \textbf{RepoShapley} & \textbf{69.51} & \textbf{88.17} & \textbf{59.14} & \textbf{84.58} \\ \midrule

\multirow{2}{*}{GLM-4.5} 
& Full-Retrieve & 65.93 & 83.04 & 55.27 & 83.21 \\
& \textbf{RepoShapley} & \textbf{69.96} & \textbf{88.49} & \textbf{60.32} & \textbf{84.94} \\ \midrule

\multirow{2}{*}{GPT-5} 
& Full-Retrieve & 66.51 & 83.73 & 55.94 & 83.82 \\
& \textbf{RepoShapley} & \textbf{70.37} & \textbf{89.08} & \textbf{60.19} & \textbf{84.41} \\ \midrule

\multirow{2}{*}{Claude Opus 4.1} 
& Full-Retrieve & 68.23 & 84.38 & 57.56 & 85.61 \\
& \textbf{RepoShapley} & \textbf{71.09} & \textbf{89.13} & \textbf{61.74} & \textbf{85.92} \\ \bottomrule

\end{tabular}
\label{tab:sota_comparison_randomized}
\end{table*}

\subsection{Interaction-Aware Proposal.}
\label{sec:ablation_delta_vs_shapley}
We examine whether gains stem solely from verification by testing \textbf{Delta+Verify}, which replaces Shapley candidates with single-chunk rankings ($\Delta_i$) under the same verifier. Table~\ref{tab:delta_ablation} shows that \textsc{RepoShapley} achieves substantially higher verified ES. This confirms that $\Delta$ scores miss \textbf{synergistic chunks} (weak in isolation), creating a hard performance ceiling. In contrast, Shapley effectively captures these high-potential interactive subsets, providing the verifier with a superior candidate pool.


\subsection{Strong Generation Models with \textsc{RepoShapley} Policy in Infilling}
\label{app:sota_infilling}

Table~\ref{tab:sota_comparison_randomized} reports results when we pair \textsc{RepoShapley} with a wide range of state-of-the-art code LMs under the infilling setting on RepoEval-Line and RepoEval-API.
For each backbone, we compare a standard \textit{Full-Retrieve} strategy against using \textsc{RepoShapley} as a selective RAG policy while keeping the same generation model.

Across all backbones, \textsc{RepoShapley} consistently improves both EM and ES on both benchmarks, indicating that coalition-aware evidence filtering is complementary to model scaling and remains effective for both open-source and closed-source generators.
For closed-source models, we use the official APIs for generation, while \textsc{RepoShapley} is applied as an external policy to decide whether to retrieve and which chunks to keep.

\subsection{Multilingual Evaluation on CrossCodeEval}
\label{app:multilingual}

To evaluate cross-lingual generalization, we extend the labeling pipeline to Java, C\#, and TypeScript, and perform multilingual mixture fine-tuning to train a single multilingual RepoShapley controller (denoted RepoShapley-M). We evaluate on CrossCodeEval~\cite{ding2023crosscodeeval} using StarCoderBase-7B and CodeLlama-13B. As shown in Table~\ref{tab:multilingual}, RepoShapley-M consistently outperforms Full-Retrieve across all four languages, with ES gains ranging from +7.4 to +10.2 points. This confirms that coalition-aware filtering is not restricted to Python and transfers effectively across languages.

\subsection{Latency-Accuracy Trade-off Analysis}

\begin{figure*}[t]
    \centering
    \includegraphics[width=1\linewidth]{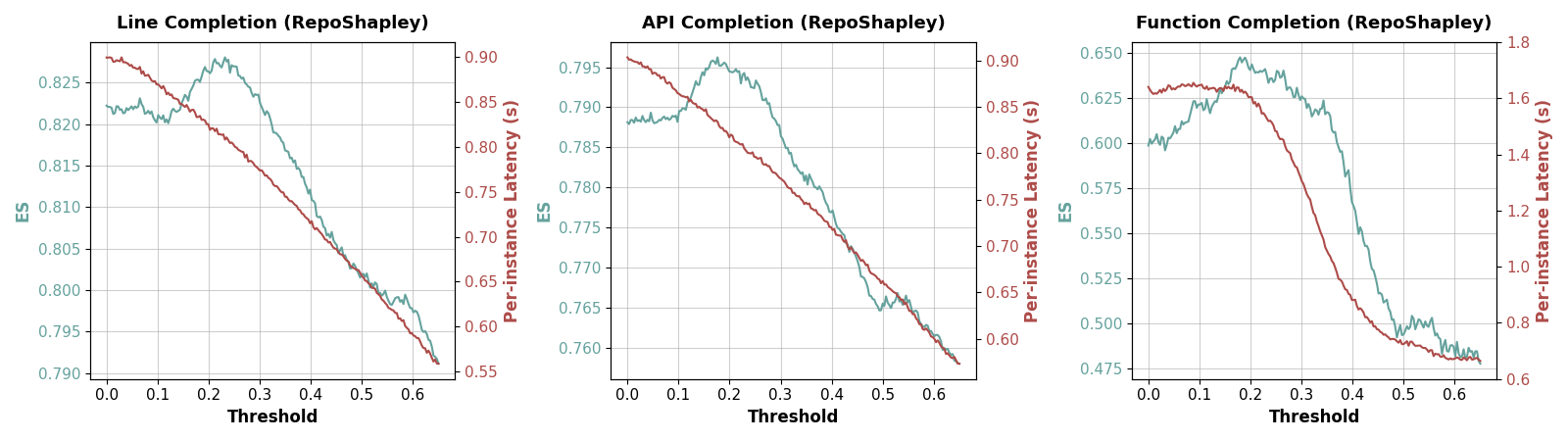}
    \caption{Latency-accuracy trade-off on SC-Base-1B.}
    \label{fig:Threshold_SC1}
\end{figure*}

\begin{figure*}[t]
    \centering
    \includegraphics[width=1\linewidth]{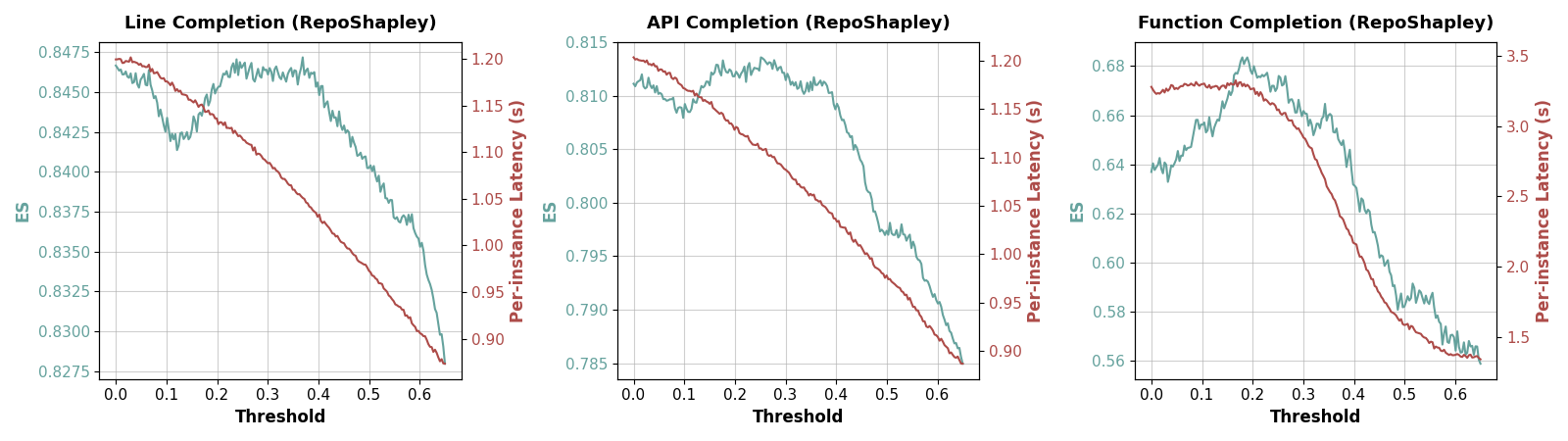}
    \caption{Latency-accuracy trade-off on SC-Base-3B.}
    \label{fig:Threshold_SC3}
\end{figure*}

\begin{figure*}[t]
    \centering
    \includegraphics[width=1\linewidth]{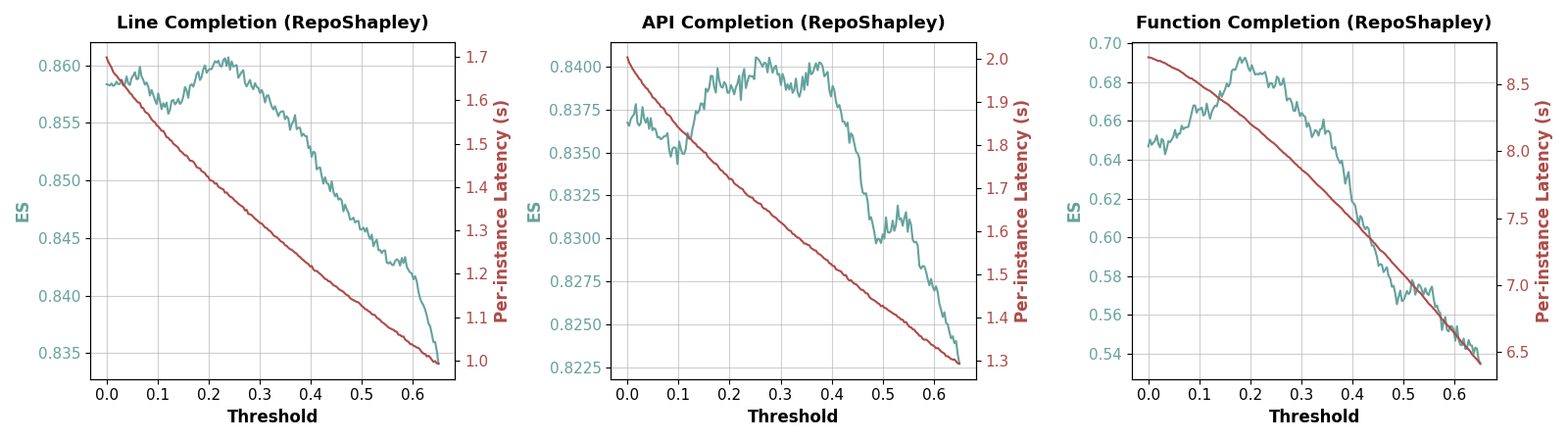}
    \caption{Latency-accuracy trade-off on SC-Base-7B.}
    \label{fig:Threshold_SC7}
\end{figure*}

Following RepoFormer~\cite{Wu2024repoformer}, we visualize the latency-accuracy trade-off to evaluate the efficiency of \textsc{RepoShapley} across StarCoderBase-1B, 3B, and 7B as shown in Figure \ref{fig:Threshold_SC1}-\ref{fig:Threshold_SC7}.
By varying the retrieval triggering threshold $t_c$ during inference, we control the model's sensitivity to external evidence.

The results demonstrate that \textsc{RepoShapley} establishes a superior Pareto frontier compared to static retrieval strategies.
We find that our model can improve accuracy while also reducing latency by skipping retrieval when the in-file context is already sufficient, and focusing retrieval on harder cases that truly need cross-file information.
Consistent with prior observations, this efficiency gain is particularly pronounced in Line and API completion tasks, where avoiding the overhead of unnecessary retrieval significantly lowers average latency without compromising generation quality.




\lstset{
    basicstyle=\ttfamily\scriptsize, 
    columns=fullflexible,
    breaklines=true,                 
    breakatwhitespace=true,          
    frame=single,                    
    rulecolor=\color{black!30},      
    showstringspaces=false,
    captionpos=b,                    
    aboveskip=5pt,
    belowskip=5pt
}

\section{Case Study}
\label{sec:case}

In this section, we present a case study to illustrate how \textsc{RepoShapley} performs interaction-aware chunk selection for repository-level code completion in the FIM setting.
The target file defines utilities for extracting event start/end markers from log search results and computing event durations.
In this instance, the missing span lies inside \texttt{LogEventStats.run}: after parsing each end-marker timestamp \texttt{end} from \texttt{end\_tag} results, the code should immediately register it into \texttt{EventCollection} via \texttt{add\_event\_end}.
After registering each end marker, the routine proceeds to handle start markers and finally calls \texttt{calculate\_event\_deltas}.
The repository contains many timestamp-related helpers; however, most retrieved evidence is only partially relevant or unrelated (e.g., YAML formatting, tests, Ceph helpers).
Naively appending all retrieved contexts can distract the model into rewriting timestamp parsing rather than emitting the required event-collection logic.

\paragraph{Instance (FIM).}
Given the in-file prefix and suffix (Figure~\ref{fig:case_fim_instance}), the model must generate the missing span at \tokmid.
Concretely, the correct completion should insert an \texttt{add\_event\_end} call that uses the current \texttt{result}'s \texttt{event\_id} and the parsed \texttt{end} timestamp.
For compact presentation, Figure~\ref{fig:case_fim_instance} splits the code across two columns.
Therefore, the \toksfx{} panel shows the subsequent lines after the insertion point (not necessarily the immediate next line in the source file), while preserving the original indentation and control flow.

\paragraph{Retrieved top-10 cross-file chunks.}
We retrieve the top-10 candidates $\{c_1,\dots,c_{10}\}$ from other files in the same repository (Figure~\ref{fig:case_chunks_select}).
Most candidates are either unrelated or only partially relevant.
Importantly, the most helpful timestamp utilities are split across multiple chunks ($c_1, c_8, c_9$), so that no single chunk alone fully specifies the needed behavior, making the evidence \emph{interaction-heavy}.

\paragraph{Why the kept coalition matters.}
Chunks $c_1$, $c_8$, and $c_9$ form a coherent timestamp-handling subroutine.
$c_1$ and $c_8$ provide compatible \texttt{datetime} parsing formats, and $c_9$ implements temporal filtering logic used by the surrounding utilities.
Full-Retrieve is distracted by irrelevant utilities (e.g., $c_2, c_3$) and hallucinates redundant timestamp-parsing logic inside \texttt{run}, instead of emitting the required \texttt{add\_event\_end} registration.

\paragraph{Generation comparison.}
Figure~\ref{fig:case_completion_compare} compares the generations.
\textsc{RepoShapley} correctly inserts the \texttt{add\_event\_end} registration and then follows the existing start-marker logic, whereas Full-Retrieve is distracted and produces redundant parsing code.

\paragraph{Failure modes.}
While \textsc{RepoShapley} improves overall performance, we observe three recurring failure patterns.
First, when all retrieved chunks are low-quality (e.g., the repository lacks relevant cross-file context), the coalition game has no good coalition to select, and the model may still keep noisy chunks rather than abstaining entirely.
Second, when chunk interactions are highly non-monotone---for example, three chunks that are individually harmful but jointly beneficial---the surrogate game's logistic form may underestimate the coalition value, causing the verifier to miss the optimal subset.
Third, in cases where the ground-truth completion depends on implicit project conventions not captured in any retrieved chunk, even a perfect coalition cannot help, and the model may hallucinate plausible but incorrect code.
We observed such failures in approximately 8\% of RepoEval instances where \textsc{RepoShapley} underperformed Full-Retrieve.

\section{Dataset Licenses and Code Availability}
\label{app:license}

\paragraph{Training data.}
Our training data is derived from the permissively licensed source-code subset of The Stack~\cite{kocetkov2023stack}. We follow the opt-out provisions specified by the dataset authors and do not redistribute raw source files.

\paragraph{Evaluation benchmarks.}
RepoEval~\cite{zhang2023repocoder} is released under the MIT License.
CrossCodeEval~\cite{ding2023crosscodeeval} is released under the Apache 2.0 License.
CrossCodeLongEval~\cite{Wu2024repoformer} follows the same license as RepoEval.

\paragraph{Models.}
StarCoderBase~\cite{li2023starcoder} is released under the BigCode OpenRAIL-M License.
CodeLlama~\cite{roziere2023codellama} is released under the Llama 2 Community License.

\paragraph{Code.}
Our code is publicly available at \url{https://github.com/yuhuo03/RepoShapley} under the Apache-2.0 License, including data preprocessing scripts, training configurations, and evaluation pipelines for reproducing all experiments.

\clearpage

\clearpage

\begin{figure*}[t]
\centering
\begin{minipage}[t]{0.48\linewidth}
\begin{casecode}[caseaccentteal]{\tokpfx}
import statistics
from datetime import datetime
class EventCollection(object):
    """Used to collect events found in logfiles..."""
    def __init__(self):
        self._events = {}
    def most_recent(self, items):
        return sorted(items, key=lambda e: e["end"], reverse=True)[0]
    @property
    def complete_events(self):
        # ... (omitted for brevity if needed) ...
        return complete
    @property
    def incomplete_events(self):
        # ... (omitted for brevity) ...
        return incomplete
    def find_most_recent_start(self, event_id, end_ts):
        """ For a given event end marker, find the most recent start marker. """
        last = None
        for item in self._events[event_id].get("heads", []):
            start_ts = item["start"]
            if start_ts <= end_ts:
                if not last or start_ts > last["start"]:
                    last = item
        return last
    def add_event_end(self, event_id, end_ts):
        if event_id not in self._events:
            self._events[event_id] = {}
        if "tails" not in self._events[event_id]:
            self._events[event_id]["tails"] = [end_ts]
        else:
            self._events[event_id]["tails"].append(end_ts)
    def add_event_start(self, event_id, start_ts, metadata=None,
                        metadata_key=None):
        # ... logic to add start markers ...
        pass 
    def calculate_event_deltas(self):
        # ... logic to calc deltas ...
        pass
\end{casecode}
\end{minipage}
\hfill
\begin{minipage}[t]{0.48\linewidth}
\begin{casecode}[caseaccentteal]{\toksfx}
class SearchResultIndices(object):
    # ... index definitions ...
    pass
class LogEventStats(object):
    """Used to identify events within logs..."""
    def __init__(self, results, results_tag_prefix, custom_idxs=None):
        self.data = EventCollection()
        self.results = results
        self.results_tag_prefix = results_tag_prefix
        # ... init logic ...
    def run(self):
        """ Collect event start/end markers... """
        seq_idxs = self.log_seq_idxs
        end_tag = "{}-end".format(self.results_tag_prefix)
        for result in self.results.find_by_tag(end_tag):
            day = result.get(seq_idxs.day)
            secs = result.get(seq_idxs.secs)
            end = "{} {}".format(day, secs)
            end = datetime.strptime(end, "%Y-%m-%d %H:%M:%S.%f")
            start = "{} {}".format(day, secs)
            start = datetime.strptime(start, "%Y-%m-%d %H:%M:%S.%f")
            metadata = result.get(seq_idxs.metadata)
            meta_key = seq_idxs.metadata_key
            event_id = result.get(seq_idxs.event_id)
            self.data.add_event_start(event_id, start, metadata=metadata,
                                      metadata_key=meta_key)
        self.data.calculate_event_deltas()
    def get_top_n_events_sorted(self, max, reverse=True):
        # ... sorting logic ...
        return top_n_sorted
    def get_event_stats(self):
        # ... stats logic ...
        return stats
\end{casecode}
\vspace{1mm}
\hfill\tokneed
\end{minipage}
\vspace{-3pt}
\caption{\textbf{FIM instance.} The missing span inserts the \texttt{add\_event\_end} registration after parsing each end marker; the subsequent start-marker handling logic continues in the suffix.}
\label{fig:case_fim_instance}
\end{figure*}

\begin{figure*}[t]
\centering
{\small\sffamily\bfseries Retrieved Candidates Pool \& Selection Decisions\par}
\vspace{2pt}

\begin{minipage}[t]{0.49\linewidth}
\begin{casecode}[casecompact,caseaccentkeep]{\tokc{1}\hspace{0.45em}\tokkeep}
ts_formats = ["%Y-%m-%d %H:%M:%S.%f", "%Y-%m-%d %H:%M:%S"]
# ... (timestamp parsing logic) ...
def filter_by_age(cls, results, result_age_hours):
    # ...
    current = datetime.strptime(current, "%Y-%m-%d %H:%M:%S")
    # ...
\end{casecode}
\vspace{0.8mm}

\begin{casecode}[casecompact,caseaccentdrop]{\tokc{2}\hspace{0.45em}\tokdrop}
if message is not None:
    message = str(message).format(**fdict)
# ... message formatting utilities ...
@cached_yproperty_attr
def type(self):
    """ Name of core.issues.IssueTypeBase object ... """
    # ...
\end{casecode}
\vspace{0.8mm}

\begin{casecode}[casecompact,caseaccentdrop]{\tokc{3}\hspace{0.45em}\tokdrop}
data_file = os.path.join(dtmp, 'data.txt')
# ... YAML generation for testing ...
class MyEventHandler(events.YEventCheckerBase):
    def __init__(self):
        super().__init__(yaml_defs_group='mygroup'...)
\end{casecode}
\vspace{0.8mm}

\begin{casecode}[casecompact,caseaccentdrop]{\tokc{4}\hspace{0.45em}\tokdrop}
def _override_keys(cls): return ['raises']
# ... YPropertyOverride logic ...
def message_with_format_dict_applied(self, property=None):
    # ...
\end{casecode}
\vspace{0.8mm}

\begin{casecode}[casecompact,caseaccentdrop]{\tokc{5}\hspace{0.45em}\tokdrop}
s = FileSearcher()
# ... Unit tests for SequenceSearchDef ...
def test_sequence_searcher_eof(self):
    # ...
\end{casecode}
\end{minipage}
\hfill
\begin{minipage}[t]{0.49\linewidth}
\begin{casecode}[casecompact,caseaccentdrop]{\tokc{6}\hspace{0.45em}\tokdrop}
# ... Duplicate or similar unit tests ...
self.assertEqual(len(sections), 2)
# ...
\end{casecode}
\vspace{0.8mm}

\begin{casecode}[casecompact,caseaccentdrop]{\tokc{7}\hspace{0.45em}\tokdrop}
""" Returns a dict of ceph versions info ... """
out = self.cli_cache['ceph_versions']
# ... Ceph daemon logic ...
\end{casecode}
\vspace{0.8mm}

\begin{casecode}[casecompact,caseaccentkeep]{\tokc{8}\hspace{0.45em}\tokkeep}
class YPropertySearchBase(YPropertyMappedOverrideBase):
    @classmethod
    def get_datetime_from_result(cls, result):
        """ This attempts to create a datetime object... """
    ts = result.get(1)
    # ...
    ts_formats = ["%Y-%m-%d %H:%M:%S.%f", ...]
    # ...
\end{casecode}
\vspace{0.8mm}

\begin{casecode}[casecompact,caseaccentkeep]{\tokc{9}\hspace{0.45em}\tokkeep}
for r in sorted(_results, key=lambda i: i[0], reverse=True):
    # ... filtering logic ...
def apply_constraints(self, results):
    # ...
    results = self.filter_by_age(results, result_age_hours)
    # ...
\end{casecode}
\vspace{0.8mm}

\begin{casecode}[casecompact,caseaccentdrop]{\tokc{10}\hspace{0.45em}\tokdrop}
relname = 'unknown'
# ... package version check ...
def _get_bind_interfaces(self, type):
    """ For the given config network type ... """
    # ...
\end{casecode}
\end{minipage}

\vspace{3pt}
\caption{\textbf{10-chunk retrieved pool and selection.} \textsc{RepoShapley} identifies and keeps $\{c_1,c_8,c_9\}$ (timestamp utilities split across chunks) while dropping unrelated evidence. (Content abbreviated for display).}
\label{fig:case_chunks_select}
\end{figure*}

\begin{figure*}[t]
\centering
\begin{minipage}[t]{0.32\linewidth}
\begin{casecode}[casecompact,caseaccentblue]{Ground Truth}
    self.data.add_event_end(
        result.get(seq_idxs.event_id),
        end)
start_tag = "{}-start".format(
    self.results_tag_prefix)
    for result in self.results.find_by_tag(
        start_tag):
    day = result.get(seq_idxs.day)
    secs = result.get(seq_idxs.secs)
\end{casecode}
\end{minipage}
\hfill
\begin{minipage}[t]{0.32\linewidth}
\begin{casecode}[casecompact,caseaccentkeep]{\textsc{RepoShapley} (Ours)}
    self.data.add_event_end(
        result.get(seq_idxs.event_id), 
        end)
start_tag = "{}-start".format(
    self.results_tag_prefix)
    for result in self.results.find_by_tag(
        start_tag):
    day = result.get(seq_idxs.day)
    secs = result.get(seq_idxs.secs)
\end{casecode}
\end{minipage}
\hfill
\begin{minipage}[t]{0.32\linewidth}
\begin{casecode}[casecompact,caseaccentdrop]{Full-Retrieve Baseline}
    ts = "{} {}".format(day, secs)
    ts_formats = ["%Y-%m-%d..."]
    for format in ts_formats:
        try:
            end = datetime.strptime(ts, format)
            break
        except ValueError:
            continue
start_tag = "{}-start".format(
    self.results_tag_prefix)
    for result in self.results.find_by_tag(
        start_tag):
    day = result.get(seq_idxs.day)
    secs = result.get(seq_idxs.secs)
\end{casecode}
\end{minipage}
\vspace{-3pt}
\caption{\textbf{Completion comparison.} \textsc{RepoShapley} inserts the correct control-flow logic. Full-Retrieve is distracted and hallucinates redundant parsing logic.}
\label{fig:case_completion_compare}
\end{figure*}

\end{document}